\documentclass[twocolumn,twocolappendix, longbib]{aastex631}

\newcommand{\fstar}{f_\star}
\newcommand{\Tobs}{T_{\rm obs}}

\newcommand{\Tcoh}{T_{\rm coh}}
\newcommand{\Sth}{S_{\rm th}}

\newcommand{\F}{2\mathcal{F}}
\def\sci#1#2{#1\times10^{#2}}
\newcommand{\FM}{2\mathcal{\widetilde{F}}}
\newcommand{\FMthresh}{2\widetilde{\mathcal{F}}_{\mathrm{thresh}}}
\newcommand{\FstatHz}{2\widetilde{\mathcal{F}}_0}
\newcommand{\Nseg}{N_{\rm seg}}
\newcommand{\mcoh}{m_{\rm coh}}
\newcommand{\msemicoh}{m_{\rm semi-coh}}

\newcommand{\Fstat}{$\mathcal{F}$-statistic\xspace}
\newcommand{\depth}{\mathcal{D}}
\newcommand{\depthc}{\mathcal{D}_{\rm Cas A}}
\newcommand{\depthv}{\mathcal{D}_{\rm Vela Jr.}}
\newcommand{\depthg}{\mathcal{D}_{\rm G347.3}}

\newcommand{\casa}{Cas~A\xspace}
\newcommand{\vela}{Vela Jr.\xspace}
\newcommand{\targetg}{G347.3$-$0.5\xspace}
\newcommand{\rmode}{$r$-mode\xspace}
\newcommand{\rmodes}{$r$-modes\xspace}
\newcommand\T{\rule{0pt}{2.6ex}}       
\newcommand\B{\rule[-1.2ex]{0pt}{0pt}} 

\def\hsens{h_{\rm sens}^{95\%}}

\newcommand{\Eq}{Eq.\@\xspace}
\newcommand{\Fig}{Fig.\@\xspace}
\newcommand{\ie}{i.e.\@\xspace}

\newcommand{\smax}{_{\textrm{max}}}
\newcommand{\smin}{_{\textrm{min}}}

\newcommand{\tage}{t_{\textrm{age}}}
\newcommand{\tagemax}{t_{\textrm{age, max}}}
\newcommand{\tagemin}{t_{\textrm{age, min}}}
\newcommand{\hmax}{h_0^{\textrm{max}}}
\newcommand{\hest}{h_0^{\textrm{est}}}
\newcommand{\hsig}{h_0^{\textrm{sig}}}
\newcommand{\Lth}{\mathcal{L}_{\textrm{th}}}

\usepackage{pgfplotstable}
\usepackage{hyperref}
\usepackage{xcolor}
\usepackage{subfigure}
\usepackage{booktabs}
\usepackage{amsmath}
\usepackage{xspace}
\usepackage{romannum}
\usepackage{comment}

\newif\ifprintauthors
\printauthorstrue  

\begin{document}
\pagenumbering{arabic}
\title{Searches for Continuous Gravitational Waves from  Supernova Remnants in the first part of the LIGO-Virgo-KAGRA Fourth Observing run}

\ifprintauthors


\author[0000-0003-4786-2698]{A.~G.~Abac}
\affiliation{Max Planck Institute for Gravitational Physics (Albert Einstein Institute), D-14476 Potsdam, Germany}
\author{I.~Abouelfettouh}
\affiliation{LIGO Hanford Observatory, Richland, WA 99352, USA}
\author{F.~Acernese}
\affiliation{Dipartimento di Farmacia, Universit\`a di Salerno, I-84084 Fisciano, Salerno, Italy}
\affiliation{INFN, Sezione di Napoli, I-80126 Napoli, Italy}
\author[0000-0002-8648-0767]{K.~Ackley}
\affiliation{University of Warwick, Coventry CV4 7AL, United Kingdom}
\author[0000-0001-5525-6255]{C.~Adamcewicz}
\affiliation{OzGrav, School of Physics \& Astronomy, Monash University, Clayton 3800, Victoria, Australia}
\author[0009-0004-2101-5428]{S.~Adhicary}
\affiliation{The Pennsylvania State University, University Park, PA 16802, USA}
\author{D.~Adhikari}
\affiliation{Max Planck Institute for Gravitational Physics (Albert Einstein Institute), D-30167 Hannover, Germany}
\affiliation{Leibniz Universit\"{a}t Hannover, D-30167 Hannover, Germany}
\author[0000-0002-4559-8427]{N.~Adhikari}
\affiliation{University of Wisconsin-Milwaukee, Milwaukee, WI 53201, USA}
\author[0000-0002-5731-5076]{R.~X.~Adhikari}
\affiliation{LIGO Laboratory, California Institute of Technology, Pasadena, CA 91125, USA}
\author{V.~K.~Adkins}
\affiliation{Louisiana State University, Baton Rouge, LA 70803, USA}
% ...

\collaboration{3000}{The LIGO Scientific Collaboration, the Virgo Collaboration, and the KAGRA Collaboration}



\else
  \author{The LIGO Scientific collaboration, the Virgo collaboration, the KAGRA collaboration}
\fi

\hypersetup{pdfauthor={NAME et al.}}

\begin{abstract}
We present results from directed searches for continuous gravitational waves from a sample of 15 nearby supernova remnants, likely hosting young neutron star candidates, using data from the first eight months of the fourth observing run (O4) of the LIGO-Virgo-KAGRA Collaboration. The analysis employs five pipelines: four semi-coherent methods -- the Band-Sampled-Data directed pipeline, Weave and two Viterbi pipelines (single- and dual-harmonic) -- and PyStoch, a cross-correlation-based pipeline. These searches cover wide frequency bands and do not assume
prior knowledge of the targets' ephemerides.
No evidence of a signal is found from any of the 15 sources. We set 95\% confidence-level upper limits on the intrinsic strain amplitude, with the most stringent constraints reaching $\sim 4 \times 10^{-26}$ near 300 Hz for the nearby source G266.2$-$1.2 (Vela Jr.). We also derive limits on neutron star ellipticity and \rmode amplitudes for the same source, with the best constraints reaching $\lesssim 10^{-7}$ and $\lesssim 10^{-5}$, respectively, at frequencies above 400 Hz. These results represent the most sensitive wide-band directed searches for continuous gravitational waves from supernova remnants to date.

\end{abstract}

\section{Introduction} 
\label{sec:intro}

Continuous gravitational waves (CWs) are long-duration, nearly monochromatic signals expected from non-axisymmetric, rapidly rotating neutron stars (NSs). While transient gravitational waves (GWs) have been directly detected from numerous compact binary mergers~\citep{GWTC-1,GWTC-2, GWTC-3, GWTC-2.1,GWTC-4_results,PhysRevD.106.043009,Nitz_2023},
CWs remain undetected. These signals arise from time-varying mass quadrupole moments caused, for example, by small deviations from perfect axial symmetry in a spinning NS.
Even minor imperfections in the star’s spherical shape can generate CWs. However, such signals are expected to be several orders of magnitude weaker in amplitude than those from mergers (see~\citealt{riles_searches_2023,wette_searches_2023,Piccinni_searches_2022} for recent reviews). 
Detecting CWs would represent a major breakthrough in astrophysics, offering direct insights into NS structure and dynamics. This includes constraints on asymmetries and crustal deformations, the equation of state of ultra-dense matter, energy loss mechanisms, and magnetic field configurations (see~e.g., \citealt{Owen:2025ata,2024CQGra..41d3001G, 2022Univ....8..619H, 2019Univ....5..217S} for reviews). 
Moreover, CW observations could uncover hidden NSs or pulsars, invisible to traditional astronomical searches, which often rely on detecting electromagnetic pulsations. A GW detection provides an independent channel to reveal these hidden objects, expanding our view of the Galactic NS population.
Though no CW detections have yet been reported, current searches are yielding significant upper limits on strain amplitude, ellipticity, and \rmode oscillation amplitudes. These limits are guided by theoretical predictions of \rmode behavior and relativistic frequency estimates from models of NSs with realistic equations of state (e.g., \citealt{2000ApJ...543..386H, Idrisy_2015}).

As such, young core-collapse supernova remnants (SNRs), known or expected to house NSs,  represent excellent targets for the search of CWs.  
In the absence of pulsations, a NS may manifest itself as a central compact object (CCO) detected in X-ray wavelengths near the center of its hosting SNR, and typified by the NS discovered by the \textit{Chandra} X-ray Observatory (CXO) in the young SNR Cassiopeia A (Cas~A) (\citealt{2000ApJ...531L..53P, 2017JPhCS.932a2006D}).
Other manifestations of young NSs include a point-like source powering a pulsar wind nebula (PWN), the synchrotron bubble inflated by the relativistic wind of the NS as it interacts with its surrounding SNR or interstellar medium, and typically observed in radio and/or X-ray energies (see e.g., \citealt{2006ARA&A..44...17G}).
Young NSs typically spin down rapidly due to strong electromagnetic and GW torques~\citep{knispel2008blandford}, complicating long-duration CW searches. 
Previous continuous GW searches for SNRs have employed various strategies with different trade-offs. Interesting sources like Cas~A and SNR~1987A were targeted for the first time in the fifth science run of initial LIGO (S5)~\citep{abadie2010first,2016PhRvD..94h2004S} while other results for SNR~1987A can be found in the more recent O2~\citep{Owen_2022} and O3~\citep{Owen_2024,Nostro} searches, which used data from the second and third Advanced LIGO-Virgo observing runs, respectively.
Other, more extensive searches, including additional targets along with Cas~A, were performed in S6~\citep{aasi2015searches} and O1-O2 data~\citep{Abbott_2019, Lindblom_2020, millhouse2020search,PhysRevD.100.024063,Papa:2020vfz,Ming_2022}. Searches in O3 data~\citep{Abbott2021_O3aSNR,O3aSNRweave,fullo3_casA}, were characterized by being the broadest and most extensive performed to date (including a set of 15 sources in~\citealt{Abbott2021_O3aSNR}), reaching sensitive results for the well known sources Cas~A and Vela Jr. in the most sensitive frequency range~\citep{fullo3_casA}. Competitive results were achieved by the latest deep searches performed using the Einstein@Home project~\citep{ming2025x,2024ApJ...977..154M,Morales_2025}.  In particular, \cite{ming2025x} reported a $\sim 27\%$ false alarm probability candidate surviving for the source G347.3–0.5, which is also analyzed in this work by three of our pipelines.  However, we did not find any compatible candidate using the search configurations adopted here.
Previous results from the same project can be found in~\cite{PhysRevD.100.024063,Papa:2020vfz,Ming_2022}. 

In this search, we apply five search pipelines to look for CWs from 15 young to middle-aged (i.e, $<$ 10$^5$~yr) SNRs using data from the O4 run of the LIGO-Virgo-KAGRA collaboration. Specifically, only the first eight months (O4a, beginning on May 24, 2023 and concluding on January 16, 2024) of data from LIGO Hanford and LIGO Livingston detectors were analyzed. All these pipelines were previously used in O3 or O3a analysis~\citep{Abbott2021_O3aSNR,O3aSNRweave,Nostro}. 
These include the directed Band-Sampled-Data (BSD) pipeline~\citep{BSD}, the single-harmonic Viterbi (SHV) and dual-harmonic Viterbi (DHV) methods~\citep{sun2018hidden,Sun2019}, and the Weave~\citep{Wette2018} semi-coherent searches, as well as the cross-correlation analysis by PyStoch~\citep{PyStoch}.
Together, the pipelines provide complementary parameter-space coverage and sensitivity to different emission scenarios.

The paper is organized as follows. Section~\ref{sec:targets} describes the SNR targets and their astrophysical properties, while in Section~\ref{sec:signal} we briefly discuss the signal models used in this search. Section~\ref{sec:methods} outlines the search methods and parameter space explored. Section~\ref{sec:results} presents the results, including upper limits and astrophysical implications.
Conclusions are summarized in Section~\ref{sec:conclusions}. Appendices~\ref{appendix:BSDmethod}--\ref{appendix:weave_pipelinesdetails} detail technical information on the pipelines, the candidate's follow-up and postprocessing, and the procedures used to set upper limits for BSD, PyStoch, SHV, DHV, and Weave, respectively.

\section{Targets}
\label{sec:targets}

\subsection{Methodology}
The SNR targets, summarized in Table~\ref{tab:targets}, are selected from the Green catalogue of radio SNRs
\citep{2025yCat.7297....0G} 
and {\tt{SNRcat}}, the online high-energy catalogue of SNRs hosted by the University of Manitoba\footnote{http://snrcat.physics.umanitoba.ca} \citep{SNRcatpaper} with their distance and age estimates regularly updated. 
The youngest objects in this sample are considered to be among the most promising candidates for CWs since their expected high spin-down implies a potentially large energy budget for radiation in GWs. 

The target selection is primarily based on a sample of core-collapse SNRs known to host isolated NSs or NS candidates. Their position has been inferred from the presence of a pulsar wind nebula or a point-like X-ray source (a CCO or CCO candidate), both of which are located within or near the center of their hosting SNRs. 
The embedded NS candidates have not yet been discovered as pulsating sources, and as such, we do not know their characteristic ages. However, their age is estimated from their hosting SNR. 

A CW search in the $\sim$10 Hz--2~kHz (the sensitive range of Earth-based interferometer detectors) would reveal indirectly the presence of objects with spin periods in the $\sim$1--200 ms range since the CW frequency is (depending on the emission mechanism) approximately twice the spin frequency of the hidden star, and does not typically exceed that value (see Sec.~\ref{sec:signal}). Such a search could help reveal the fastest and youngest spinning NSs. This range of rotation period values overlaps with the known pulsation periods of many young rotation-powered pulsars as well as that of the 3 known pulsars in the CCOs family (with spin periods of 105, 112, and 424 ms; \citealt{2013ApJ...765...58G}).

\subsection{Target list and comparison to O3a}
The list of targeted SNRs is essentially the same as that chosen for O3a \citep{Abbott2021_O3aSNR} with some updates summarized next.
Firstly, SNR~1987A (G279.7$-$31.9) was added given it is believed to have now entered the remnant stage (i.e. the supernova blast wave is interacting with the circumstellar medium, \citealt{2024ApJ...966..147R}) and shows evidence of a long sought compact object or another powering engine \citep{2024Sci...383..898F, 2021ApJ...908L..45G}. This is also the youngest and only target in our sample with a well-known age and distance (being located in the Large Magellanic Cloud). 

The previously considered target, G354.4+0.0 (also known as G354.46+0.07), is removed from our O4a list since it is no longer believed to be a SNR, but rather an HII region \citep{2019PASA...36...48H}. 
G1.9+0.3 is kept in our O4a list despite it being commonly believed to be a type Ia remnant \citep{2013ApJ...771L...9B}. This is because a core-collapse origin is not completely ruled out, given the SNR's location close to the Galactic Center (e.g., \citealt{reynolds2008youngest})
and the known uncertainties in typing young SNRs from X-ray ejecta fits to nucleosynthesis yields
(e.g., \citealt{2013ApJ...771L...9B, 2023MNRAS.525.6257B, 2026ApJ...997..211T}). 
Being the youngest known Galactic SNR, G1.9+0.3 is potentially a promising target for CWs search which may provide new insight into its nature.

Furthermore, the age and distance estimates shown in Table~\ref{tab:targets} have been updated since O3a to reflect the values in {\tt{SNRcat}} (2025 version), including the most recent studies of these objects. We note that for Vela Jr., our distance upper limit is consistent with the most recent accurate distance measurement of 1.41$\pm$0.14~kpc \citep{2026ApJ...997L..36S}. These estimates affect the search parameters differently for each pipeline, as detailed in the following sections. 

Of the 15 SNRs in Table~\ref{tab:targets}, 
at least eight targets are searched using three different pipelines, while the remaining SNRs are searched using one or two pipelines.
This is because each pipeline covers different parts of the parameter space, corresponding to slightly different assumptions of the characteristic age of the system. Furthermore, two remnants, G291.0$-$0.1 (MSH~11--62) and G330.2+1.0, already searched for in O3a by the SHV method (see Sec.~\ref{sec:shv}), are added as targets in the BSD analysis (Sec.~\ref{sec:BSD}). G291.0$-$0.1 is a plerionic composite (i.e. a composite type SNR harboring a pulsar wind nebula) powered by a pulsar candidate (CXOU J111148.6$-$603926), and G330.2+1.0 hosts an X-ray emitting CCO (CXOU J160103.1$-$513353). These are young to middle-aged remnants of ages 1.3--10~kyr and 1--15~kyr, respectively.

\begin{table*}[!tbh]
\centering
\setlength{\tabcolsep}{3.5pt} 
\renewcommand\arraystretch{1.2} 
\resizebox{\textwidth}{!}{ 
\begin{tabular}{lllllll}
\toprule
Source & Compact Object or PWN association& Age& Distance & Right ascension & Declination & Pipeline \\
       &    & (kyr) & (kpc) & (h:m:s) & ($^\circ$:$'$:$''$)& \\
\toprule
G1.9+0.3 & -- & 0.15--0.22    & 8.34--8.5   & 17:48:46.9  & $-$27:10:16 & SHV \\
\hline
G15.9+0.2              & CCO CXOU J181852.0$-$150213 & 1--3    & 5--16 & 18:18:52.1  & $-$15:02:14 & SHV\\
G18.9$-$1.1 G18.9-1.1 G18.9--1.1& PWN, CXOU J182913.1$-$125113 & 2.8--6.1    & 1.6--4.7    & 18:29:13.1 & $-$12:51:13  & DHV, SHV\\ 
G39.2$-$0.3/3C 396 & PWN, CXOU J190404.7+052711.8 & 4.7--5.2 & 6.2--9.5  & 19:04:04.7 & 5:27:12 &  BSD, DHV, SHV     \\ 
G65.7+1.2/DA 495  & PWN, CXO J195217.04 +292552.5 & 3--155   & 1--3.6  & 19:52:17.0 & 29:25:53        &  BSD, DHV, SHV  \\
G93.3+6.9/DA 530  & PWN candidate, CXOU J205214+551722 & 3.2--7  & 4.4--8.3 & 20:52:14.0 & 55:17:22        & BSD, DHV, SHV  \\
G111.7$-$2.1/Cas~A      & CCO CXOUJ232327.9+584842 & 0.32--0.35  & 3.3--3.7  & 23:23:27.9  & 58:48:42  & SHV, PyStoch, Weave \\
G189.1+3.0/IC 443 & PWN G189.23+2.90 & 3--30  & 0.7--2.0  & 06:17:05.3 & 22:21:27 & BSD, DHV, SHV    \\ 
G266.2$-$1.2/Vela Jr. &  CCO CXOU J085201.4$-$461753 & 2.4--5.1 & 0.5--1.3 & 08:52:01.4 & $-$46:17:53&  BSD, DHV, SHV\\
 &  &  &  &   & & PyStoch, Weave \\
G291.0$-$0.1/MSH 11--62 & PWN G291.02$-$0.11, CXOU J111148.6$-$603926 & 1.3--10      & 3--10   & 11:11:48.6  & $-$60:39:26 & BSD, SHV\\
G330.2+1.0             &  CCO CXOU J160103.1$-$513353 & 1--15     &   5--10  & 16:01:03.1 & $-$51:33:5 & BSD, SHV \\
G347.3$-$0.5/RX~J1713.7-3946           & CCO 1WGA J1713.4-3949 &  1--1.7     & 0.9--1.1  & 17:13:28.3 & $-$39:49:53 & SHV, PyStoch, Weave\\
G350.1$-$0.3            & CCO candidate XMMU J172054.5-372652 & $\leq$0.7 & 4.5--9  & 17:20:54.5 & $-$37:26:52     &  SHV \\
G353.6$-$0.7 & CCO XMMU J173203.3-344518 & 2.4--27    & 2.4--6.1  & 17:32:03.3 & $-$34:45:18 & BSD, DHV, SHV  \\  
\hline
G279.7$-$31.9/SNR~1987A  &  NS or PWN & 0.038      & 50 & 05:35:28.0 & $-$69:16:13 &   PyStoch \\
\bottomrule
\end{tabular}
}
\caption{\label{tab:targets}
The 15 SNRs covered in this analysis, listed in their ascending order of Galactic longitude (for the Galactic remnants), and with any associated Central Compact Object (CCO) or pulsar wind nebula (PWN). 
All are Galactic SNRs except for the last source, SNR~1987A, in the Large Magellanic Cloud.  The first source G1.9+0.3 is commonly believed to be a Type Ia SNR, however, it was included given the uncertainties in SNR classification and given it is the youngest known SNR in the Milky Way galaxy.
RA and Dec are in J2000 coordinates. CXOU (XMMU) refers to J2000 coordinates of the associated NS candidate source detected in X-rays with Chandra (XMM-Newton). All 5 pipelines were applied to the Vela Jr. SNR (G266.2$-$1.2); whereas for the other SNRs, only selected pipelines were applied. 
See Section~\ref{sec:targets} for details.}
\end{table*}

\section{Signal and Emission models}
\label{sec:signal}
A rotating NS with non-zero equatorial ellipticity has a time-varying mass quadrupole moment and emits CWs. The amplitude of this emission is determined by the star’s internal structure, including factors such as the nuclear equation of state, crustal rigidity, and magnetic field configuration~\citep{2018ASSL..457..673G}.
Young NSs, in particular, are expected to retain relatively large deformations acquired at birth~\citep{knispel2008blandford}, although these asymmetries typically decay over time due to thermal, magnetic, and viscous relaxation processes~\citep{haensel1990ohmic,gnedin2001thermal,potekhin2015neutron}.

As the NS spins, it gradually loses angular momentum through several mechanisms, including GW emission. This energy loss causes a reduction in its spin (and GW) frequency, which can be modelled as \footnote{Given the proportionality between the GW frequency and the star's spin frequency $f_{\star}$, we use $f$ to denote either quantity in the formulae above. We will specify the relation between the GW frequency and $f_{\star}$ whenever it is needed.}
\begin{equation}
\label{eq:powlaw}
    \dot{f} \propto -f^{n}, ~n = \frac{f \ddot{f}}{\dot{f}^{2}}
    \end{equation}
where the \emph{braking index} $n$, is a dimensionless parameter that describes the physical mechanism driving the neutron star's energy loss. 
In practice, the spin-down evolution may be more complex than the simple power-law behaviour described by Eq.~\eqref{eq:powlaw}. The total energy loss may involve simultaneous contributions from both gravitational and electromagnetic (e.g., magnetic dipole) radiation. Some of the pipelines in this search look for signals with braking indices in the 2--7 range as in~\cite{abadie2010first} where the lowest $n=2$ value is compatible with the braking indices of
nearly all known pulsars~\citep{Hamil_2015}. In the case of spin-down dominated entirely by mass quadrupoles GW emission (as in the so-called \emph{gravitar} scenario) $n = 5$, or $n = 7$ if the emission arises from $r$-mode oscillations~\citep{Palomba2005}. For magnetic dipole radiation, the braking index takes the value $n = 3$.
The age of the NS can be estimated as
\begin{equation}
\tage = -\frac{f}{(n-1) \dot{f}} \left[1 - \left(\frac{f}{f_{\mathrm{i}}}\right)^{n-1} \right] 
\underset{f \ll f_{\mathrm{i}}}{\approx} -\frac{f}{(n-1)\dot{f}},
\label{Eq:age}
\end{equation}
where $f_{\mathrm{i}}$ is the spin frequency at birth. For older sources, where $f \ll f_{\mathrm{i}}$, the expression is simply a function of $f$ and $\dot{f}$ for a fixed braking index $n$.

The expected GW frequency depends on the emission mechanism and the star's spinning frequency $f_\star$: thermoelastic or magnetic quadrupoles emit at $f_\star$ or $2f_\star$ \citep{ushomirsky2000deformations,cutler2002gravitational,Lasky2013}, while $r$-modes emit at approximately $4f_\star/3$ \citep{Idrisy_2015,Gittins:2022rxs,1998PhRvD..58h4020O,1998ApJ...502..708A,caride2019search}. Additional contributions may arise from superfluid dynamics within the NS, potentially leading to dual-harmonic emission scenarios ($f_\star+2f_\star$) \citep{Jones2010} or even to emission at higher-order harmonics of the spin frequency~\citep{2015ApJ...807..132M,cheunchitra_persistent_2024}.
The dual-harmonic signal model, with components at $f_\star$ and $2f_\star$, is given by \citep{Jaranowski1998,Sun2019}
\begin{eqnarray}
\label{eqn:h2p} 
h_{2+}(t) &=& \frac{1}{2}h_0(1+\cos^2\iota)\sin^2\theta\cos \Phi (t),\\
h_{2\times}(t) &=& h_0 \cos \iota \sin^2\theta \sin \Phi (t), \\
h_{1+}(t)&=&\frac{1}{8}h_0\sin 2\iota \sin 2\theta \sin \frac{\Phi (t)}{2},\\ 
\label{eqn:h1c}
h_{1\times}(t) &=& \frac{1}{4} h_0 \sin \iota \sin 2 \theta \cos \frac{\Phi (t)}{2},
\end{eqnarray}
where $\iota$ is the source inclination angle, $\theta$ is the wobble angle between the star's rotation axis and its principal axis of the moment of inertia, and $\Phi(t)$ is the GW phase (at $2f_\star$) observed at the detector.\footnote{Here, we do not consider the scenarios with emissions at additional frequencies, e.g., when precession and triaxiality of the star are also considered~\citep{Zimmermann1979,VanDenBroeck2005,Lasky2013}.}
Standard CW searches target the simpler ``triaxial star, principal-axis'' model ($\theta=\pi/2$), where the $h_1$ terms vanish, leaving only the $h_2$ component at $2f_\star$. The measured detector strain is then $h(t)=F_+(t;\alpha,\delta,\psi)h_{2+}(t)+F_\times(t;\alpha,\delta,\psi)h_{2\times}(t)$, with $F_+,F_\times$ the antenna patterns for sky location $(\alpha,\delta)$ and polarization angle $\psi$.
The phase of the GW signal is primarily determined by the intrinsic evolution of the source’s spin frequency, characterized by its frequency and derivatives \((f, \dot{f}, \ddot{f}, \ldots)\) at a chosen reference time \(t_0\) (with respect to the Solar System Barycenter):

\begin{equation}
\label{eq:phase}
\begin{aligned}
\Phi(t) = \phi_0 + 2\pi \Big[ 
& f(t - t_0) 
+ \frac{\dot{f}}{2}(t - t_0)^2  \\
& + \frac{\ddot{f}}{6}(t - t_0)^3 
+ \dots 
\Big].
\end{aligned}
\end{equation}

where \(\phi_0\) is a constant phase.   
In addition to this intrinsic evolution, the observed phase is modulated by Doppler variations due to the relative motion between the source and the detector, as well as by the relativistic Einstein and Shapiro delays. In this search, we do not explicitly account for glitches; we assume that none of the targets have glitched during the considered observing time.
We note that in addition to secular spin-down, random spin variations (timing noise or spin wandering) can affect the signal phase evolution~\citep{10.1093/mnras/stz2383,namkham2019diagnostics,10.1093/mnras/staa615}, driven by complex internal and magnetospheric processes ~\citep{1981ApJ...245.1060C,melatos2014pulsar}.
These complications motivate the use of semi-coherent methods (see Sec. \ref{sec:methods}), which coherently integrate short data segments (of duration $T_{\rm coh}$) and combine the results incoherently across the full observing span.
In the absence of electromagnetic measurements of $(f,\dot{f},\ldots)$, the maximum strain $\hmax$ emitted by a rotating NS (assuming its spin-down is fully dominated by GW emission) is inferred from the indirect, age-based spin-down limit for a source located at a distance  $D$ \citep{wette_searching_2008}:

\begin{equation}
\label{eq:hmax}
h_0^{\rm max} = 2.27\times 10^{-24} 
\Bigg(\frac{1 \, {\rm kpc}}{D}\Bigg)
\Bigg(\frac{1 \, {\rm kyr}}{t_{\rm age} }\frac{I_{zz}}{10^{38} \, {\rm kg \, m}^2}\Bigg)^{1/2}
\end{equation}
where  $I_{zz}$ is the NS's moment of inertia.

\section{Methods}
\label{sec:methods}
\begin{figure}[tbh]
    \centering
    \includegraphics[width=1\linewidth]{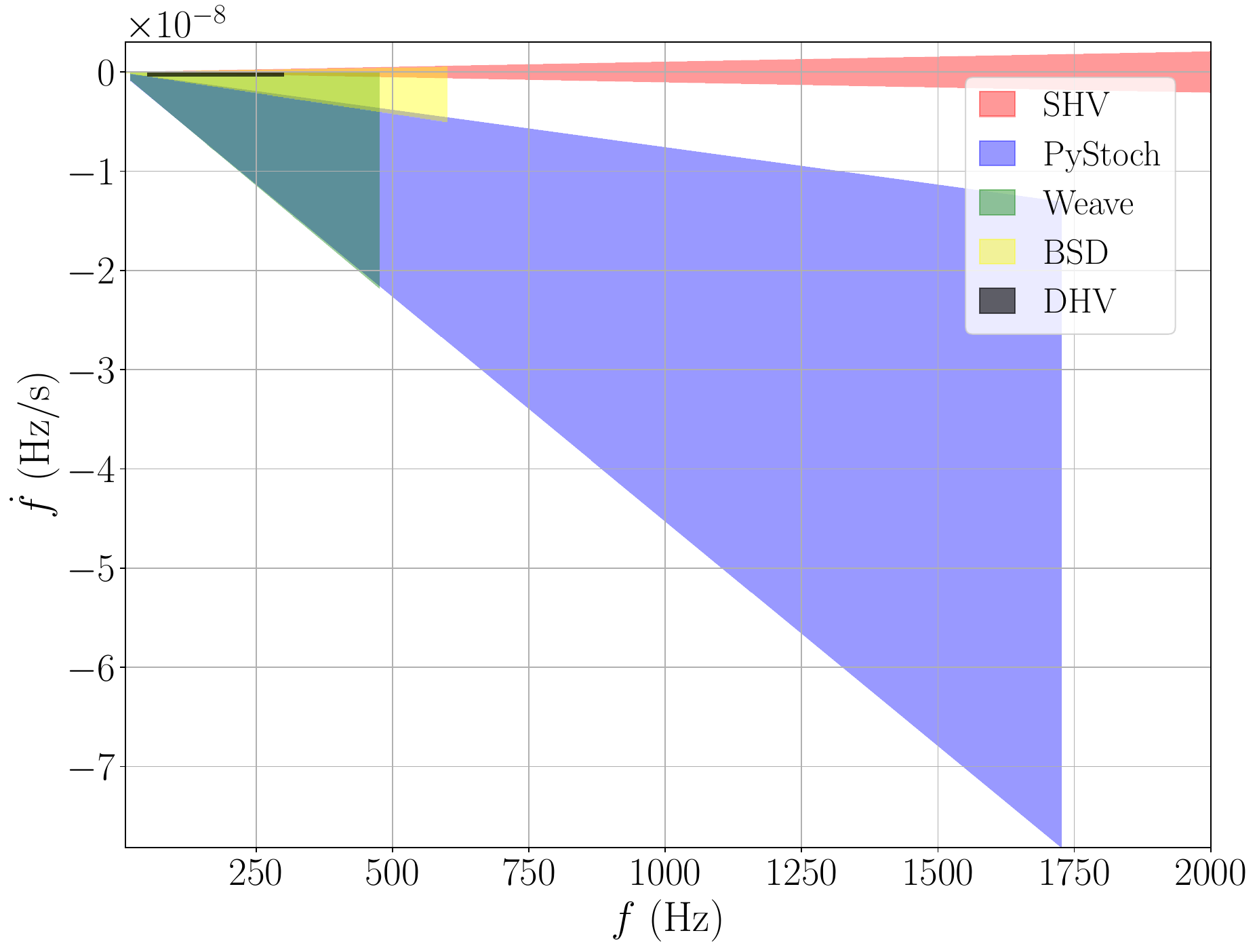}
    \caption{Parameter space covered by each of the 5 pipelines for G266.2$-$1.2 (Vela Jr.). } 
    \label{fig:parspacevelajr}
\end{figure}

This section briefly describes each of the five pipelines applied in this search, the parameter space investigated and the emission mechanism considered. 
Pipeline details, along with the description of the post-processing of the selected candidates, are reported in the Appendices. 
Each pipeline has a different approach when selecting the parameter space for the search.
For comparison, we report in Fig.~\ref {fig:parspacevelajr} the values of the frequency $f$ and first-order spin-down/up $\dot{f}$ parameter spanned by each search for Vela Jr., the only source targeted by all five pipelines. The different $(f, \dot{f})$ ranges covered by each pipeline arise from the distinct choices of signal emission models and from the computational constraints of each method, as described in Secs.~\ref{sec:BSD}--\ref{sec:weave method}.
PyStoch and Weave also include the second-order spin-down parameter in the signal template as indicated in Appendix ~\ref{appendix:pystochsearchtargets} and ~\ref{appendix:weave_searchtargets}, respectively.

\subsection{Data}
All the pipelines use data from the LIGO Hanford and Livingston detectors~\citep{LIGO2014} in this search. The analysis covers the data collected between May 24, 2023, 15:00 UTC and January 16, 2024, 16:00 UTC -- the period referred to as O4a~\citep{Jia_2024,PhysRevX.13.041021, PhysRevD.111.062002,2025arXiv250818079T,Soni_2025}. All the pipelines use data collected when the detectors are in the science observing mode~\citep{science_mode_O4a}. The low-latency, cleaned, analysis-ready, calibrated data -- C00 frames
(channel names: \texttt{H1:GDS-CALIB\_STRAIN\_CLEAN\_AR} and \texttt{L1:GDS-CALIB\_STRAIN\_CLEAN\_AR})
~\citep{Wade_2025,O4cal, Sun_2020,Viets2018,Viets_thesis,Vajente2020} have been used to generate the different input files for each pipeline. All the inputs have passed through a glitch-gating procedure~\citep{o4gating, BSD, Acernese_2009}. The time-domain BSD files, which are generated from the Short FFT Database (SFDB) files~\citep{FFTpeakmaps}, a collection of short-time Fourier Transform of duration 1024 s, have been cleaned using a double-gating procedure described in~\cite{BSD,Acernese_2009}.
PyStoch begins its analysis using folded Stochastic Intermediate Data (SID), which are ultimately derived from the frames files. The SID consist of cross-correlated spectra and cross-detector noise estimates computed with a coherence time of 192 s and frequency resolution of 1/32 Hz~\citep{pygwb_paper}. These data are then folded into a single sidereal day, resulting in the \emph{folded} SID~\citep{PyStoch}.  
The remaining pipelines start their analyses from the 30-minute Short Fourier Transforms (SFTs)~\citep{LSC_sft_2022}. The SHV and DHV pipelines used the G02 version of gated data~\citep{o4gating} while the Weave method used the earlier G01 release, which had spurious gates near boundaries of buffered data segments, a bug found to have negligible influence on search sensitivity. The lists of instrumental artefacts used to exclude some contaminated frequency bands are reported in~\cite{known_lines_O4a}.

\subsection{Band-Sampled-Data (BSD)}
\label{sec:BSD}
The BSD pipeline, adapted for the SNR case~\citep{BSD,Abbott2021_O3aSNR}, is a directed semi-coherent search method based on the Frequency-Hough (FH) transform~\citep{FrequencyHoughmethod, 2008CQGra..25r4015A}.
Previous applications of this pipeline targeting the Galactic Center during O2 and O3 are reported in \cite{DirectedBSD,O3GC}, respectively. 
In this analysis, we target a subset of six sources, already searched for by the same pipeline in O3a~\citep{Abbott2021_O3aSNR}, with the addition of two new targets: G291.0$-$0.1 and G330.2+1.0. The full pipeline details can be found in~\cite{Abbott2021_O3aSNR} and references therein.
The signal model used in this search neglects the contributions from the second-order spin-down parameter (assumed always to be contained in a single $\ddot{f_0}$ bin).
The first order spin-down range is defined as $-f_{\rm max}/\tage \leq \dot{f} \leq 0.1 f_{\rm max}/\tage$, with $f_{\rm max}$ the upper frequency of each 10-Hz band.
We search both negative and positive spin-down to allow for possible spin-up scenarios. 
When available, we use the youngest reliable age estimate for each remnant from {\tt{SNRcat}}'s 2025 version \citep{SNRcatpaper}.
The source age used in the search also determines the range of second-order spin-down values that must be covered. However, this range will be larger for higher frequencies, determining de facto a limit on the maximum frequency used in this search (See Appendix~\ref{appendix:BSDmethod}). Conversely, if the maximum frequency is fixed, the resulting second-order spin-down range places a lower bound on the source age (see e.g. Eq (B2) of~\citealt{Abbott2021_O3aSNR}). In this O4a analysis, we set a maximum frequency of 600~Hz, which corresponds to a minimum source age of $3.8~$kyr.  
The list of targets and corresponding parameter space configurations is summarized in Appendix~\ref{appendix:BSDserchsetuptargets}.

\subsection{PyStoch}
The PyStoch pipeline \citep{PyStoch} employs a Python-based tool built on the radiometer method, typically used to map the Stochastic GW Background across the sky. It analyzes cross-correlated and folded data from the LIGO Hanford and LIGO Livingston detectors 
in the [20–1726]~Hz band. A previous search using O3 data can be found in~\cite{Nostro}. In this search, PyStoch estimates the optimal signal-to-noise ratio as a function of frequency for a specific sky location.
This study focuses on four well-known SNRs: SNR~1987A, Vela Jr., Cas~A, and G347.3$-$0.5; see Table~\ref{tab:targets}. The explored parameter space for each target is summarized in Appendix~\ref{appendix:pystochsearchtargets}. The lower and upper bounds delineate the limiting cases of the physical scenarios being analyzed, i.e.: 
\begin{equation}
\label{eq:PyStoch_ranges}
  -\frac{f}{\tage} \leq \dot{f} \leq -\frac{f}{6\tage}, \quad
  2\frac{\dot{f}^2}{f} \leq \ddot{f} \leq 7\frac{\dot{f}^2}{f}
\end{equation}
including the frequency distribution parameter in Eq.~\eqref{Deltaf2}.

\subsection{Viterbi (single-harmonic)}
\label{sec:shv}
The SHV method implements a hidden Markov model (HMM) to efficiently track CW signals from NSs exhibiting secular spin down and spin wandering \citep{suvorova_hidden_2016,sun2018hidden}. 
Previous searches targeting young SNRs using this algorithm were conducted on Advanced LIGO's data from the second observing run (O2)\citep{millhouse2020search,Beniwal2022} and the first half of the third LIGO-Virgo-KAGRA observing run (O3a) \citep{Abbott2021_O3aSNR}.
In this analysis, we search for 14 targets; 
For a given SNR with distance $D_{\textrm{min}} \leq D \leq D\smax$ and age $\tagemin \leq \tage \leq \tagemax$, we define the searchable parameter space according to four criteria described in detail in Appendix~\ref{appendix:SHV_searchtargets}.

\subsection{Viterbi (dual-harmonic)}
\label{appendix:dhv}
The DHV method~\citep{Sun2019} is a semi-coherent tracking algorithm designed to follow two frequency components simultaneously, improving sensitivity to signals emitted at both the stellar rotation frequency $\fstar$ and its second harmonic, $2\fstar$. Other methods in this paper assume the conventional signal model in which the GWs are emitted only at a frequency twice the spin frequency of the star, $2 \fstar$.
In addition, we assume that the signal frequency evolution is dominated by the star's secular spin-down and approximated by a negatively biased random walk. The star's unknown spin-down rate, which can vary over time, is covered in a range between zero and the maximum allowed spin-down rate $\dot{f}_{\star_{\rm max}}$, determined by the search configuration (see Appendix~\ref{appendix:dhv_details}).
The list of targets and parameter space covered are given in Appendix~\ref{appendix:dhv_detailssearchtargets} . 
Two search configurations are used for relatively older and younger targets. 
A minimum assumed characteristic age of the target, $\tage$, is selected to be either 20~kyr or 5~kyr, based on the estimated ages of the remnants (see Table~\ref{tab:targets}). 
To mitigate the impact of noise artifacts, particularly due to combining dual harmonics, we impose a lower frequency cutoff at 50~Hz for $\fstar$ in all targets. This avoids contamination from instrumental lines that are more prevalent at lower frequencies. 
The maximum frequency is set by the selected $\tage$, i.e.,
${f_\star}_{\rm max}=|\dot{f}_{\star_{\rm max}}|(n-1) \tage$. Assuming that the spin down is dominated by gravitational radiation, we take $n=5$~\citep{sun2018hidden,Abbott2021_O3aSNR}.

\subsection{Weave}
\label{sec:weave method}
The Weave pipeline is a template-based, semi-coherent search method based on the \Fstat~\citep{Jaranowski1998}. 
The \Fstat computed for each segment analyzed is proportional to the square of the signal amplitude $h_0^2$, maximized over $h_0$, the unknown orientation angles $\iota$ and $\psi$, and the phase constant $\phi_0$. 
Thus, the search focuses on the parameters governing the phase evolution in Eq.~\eqref{eq:phase}.
Due to the high computational cost of template-based Weave searches, we limit our search to \casa, \vela, and \targetg for GW signal frequency $f$ ranges from 20 to 475 Hz. 
The search ranges for frequency derivatives are defined by assuming a power-law spin-down: $\dot{f} \propto -f^n$, with a braking index $n$ ranging between 2 and 7. 
The maximum absolute values of $\dot{f}$ and $\ddot{f}$ at the lowest and highest search frequencies are reported in Appendix~\ref{appendix:weave_searchtargets}.

\section{Results}
\label{sec:results} 
We do not find any statistically significant candidate from all our targets. All the candidates from each pipeline were discarded after the post-processing and follow-up procedures described in Appendices~\ref{appendix:BSDpostprocessingFU}--\ref{appendix: weave follow-up}. None of the candidates surviving prior to the final steps and found with the reported search configurations seems to resemble the one listed in~\cite{ming2025x}. 
This may be expected, given the modest false alarm probability of ~27\% reported for the candidate.
Nevertheless, to fully assess the nature of this candidate, more investigations using a fully coherent method on the full O4 dataset are needed.

In each case, the most significant outliers are consistent with instrumental artifacts or noise fluctuations. We report upper limits on the GW strain amplitude $h_0$ for each target and pipeline, including both single- and dual-harmonic models where applicable. These results are summarized in Sec~\ref{sec:ULs}, and demonstrate a marked improvement over previous observing runs, particularly in frequency bands where the sensitivity of the detectors was optimal. For several remnants, the achieved upper limits surpass indirect spin-down constraints over wide frequency ranges. As discussed in Sec.~\ref{sec:astrolimits}, the upper limit curves are translated into the ellipticity (i.e. the maximum degree of deformation the star can sustain) and, for the case of an emission happening via \rmodes, into the upper bound on the \rmode amplitude parameter. Despite the non-detection, these results contribute to valuable constraints by placing meaningful bounds on the ellipticity and \rmode amplitude of the putative NSs. 

\subsection{Upper limits}
\label{sec:ULs}
In the following, we report the $95\%$ confidence level upper limits for each pipeline. 
These limits are obtained using different approaches, including injection-and-recovery methods or semi-analytic estimates that were spot-checked against injections (see Appendices~\ref{appendix:BSDUL}--\ref{appendix: weave ul injections}).
A common figure of merit for quantifying search sensitivity is known as the sensitivity depth $\depth$~\citep{Behnke2015,PhysRevD.98.084058}:
\begin{equation}
  \depth(f) \equiv \frac{\sqrt{S_h(f)}}{h_0^{95\%}},
  \label{eqn:sensitivitydepth}
\end{equation}
\noindent where $\sqrt{S_h(f)}$ is an estimate of the effective strain amplitude spectral noise density. This quantity is independently estimated for each pipeline, as the input data used may differ between methods. We quantify the sensitivity depth for each pipeline. 

\begin{figure*}[tbh]
    \centering
    \includegraphics[width=1\linewidth]{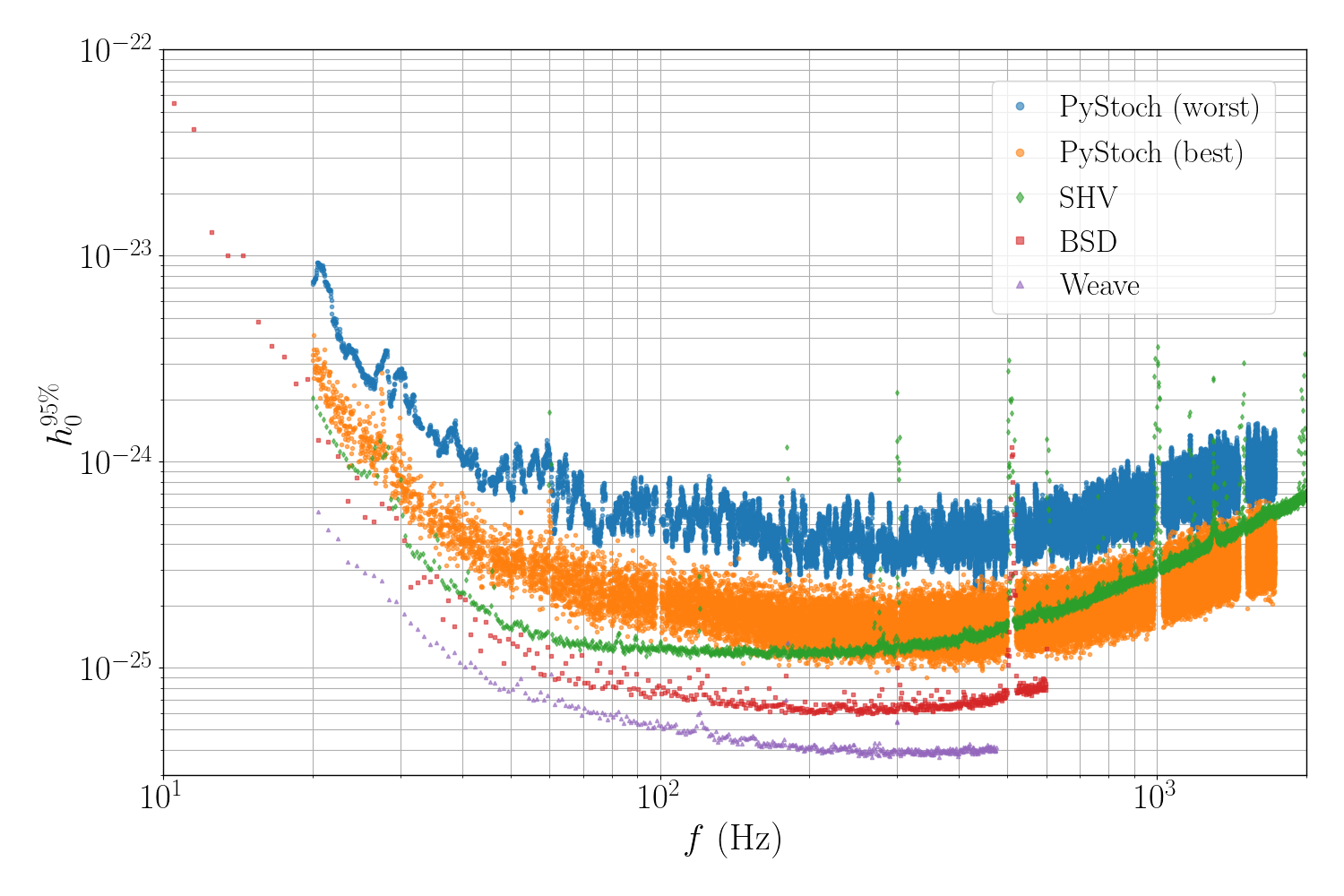}
    \caption{Upper Limits for the source G266.2$-$1.2 (Vela Jr.) obtained for the pipelines assuming the single harmonic emission scenario. The dual harmonic results for the same source are shown in Fig.~\ref{fig:viterbi1f2fUL}.}.
    \label{fig:UL_velajrall}
\end{figure*}

In the following, we briefly describe the methodology used to set these upper limits in each pipeline. As a reference, in Fig.~\ref{fig:UL_velajrall} we report the results for Vela Jr., the only common target among all the pipelines. We note, however, that DHV results are not reported, as the interpretation of these upper limits for a dual-harmonic scenario is not directly comparable with that of a single-harmonic emission model. The upper limit from the sources investigated by DHV can be found in Fig.~\ref{fig:viterbi1f2fUL}, where the best Vela Jr. upper limit is $2.12\times{10^{-25}}$ reached at 280 Hz. \\

\emph{BSD results:}
All the candidates surviving from the BSD pipeline are consistent with noise fluctuations (See 
Appendix ~\ref{appendix:BSDpostprocessingFU}), and no evidence of a CW is found in the LIGO O4a dataset. Constraints on the GW strain amplitude are derived using a semi-analytical approach previously applied in~\cite{DirectedBSD,O3GC}. 
We corroborate the results by computing the $95\%$ confidence level upper limits on $h_0$ using artificial signals injected in randomly selected 1-Hz sub-bands.
Results for all the targets are reported in  
Appendix~\ref{appendix:BSDUL}.
For all the analysed targets, the best strain sensitivity exceeds the indirect limits even when assuming the worst-case scenario (i.e., a very old source at the highest distance, as shown in Table~\ref{tab:targets}), except for G65.7+1.2 and G330.2+1.0. When considering the best-case scenario, corresponding to young and nearby sources, all the targets exceed the indirect age-based limit for frequencies above 50 Hz, except in highly contaminated frequency bands.

The best achieved results at $95\%$ confidence level is $h_0 = 5.94 \times 10^{-26}$ at around 250 Hz for the nearby source Vela Jr. The sensitivity depth for Vela Jr. at 200 Hz is 59.1 $\text{Hz}^{-1/2}$, while a maximum value of 81.8~$\text{Hz}^{-1/2}$ is reached for the lowest frequency band for the same source as expected by the use of a longer coherence time. The lowest sensitivity depth among all targets at 200 Hz is achieved for G39.2$-$0.3 with 45.8 $\text{Hz}^{-1/2}$.\\

\emph{PyStoch results:} None of the candidates survived the two follow-up procedures described in Appendix~\ref{appendix:PystochFU}. Subsequently, 95\% confidence-level upper limits on the strain amplitude are computed across a range of frequency resolutions. The computation of the upper limits is performed within a Bayesian framework, incorporating prior distributions over relevant source parameters such as the inclination angle, polarization angle, and calibration uncertainty. The final result is a marginalized posterior distribution for $h_0$, from which the upper limit at a given confidence level (e.g., 95\%) is extracted.
The most stringent upper limits are obtained using the default resolution of $1/32$~Hz.
The results for the whole frequency band are reported in Appendix~\ref{appendix:pystochUL}. 
The best upper limits are obtained using the default resolution, across the 20--1726 Hz band for each target in O4a. 
As expected, the search sensitivity degrades with increasing frequency bin combination: the results obtained directly from the default bin resolution yield tighter upper limits, while broader bin combinations are crucial to improve the signal-to-noise ratio of a signal spread over a broader frequency interval. The latter case, due to error propagation, will be still be worse than the case of a narrower signal.
The best upper limit is achieved for Vela Jr. at $\sim320$ Hz, reaching $9.02\times 10^{-26}$. We update the best upper limit on the SNR~1987A reaching $9.83\times 10^{-26}$, beating previous results of~\cite{Owen_2024,Nostro}. The sensitivity depth for Vela Jr. at the best (worst) frequency resolution is 36.5 (14.9) $\text{Hz}^{-1/2}$. The best sensitivity depth at the best frequency resolution is achieved for Cas~A at 1693 Hz with 41.6 $\text{Hz}^{-1/2}$  and for G347.3$-$0.5 at 1161 Hz with 16.4 $\text{Hz}^{-1/2}$, with the latest being too low to reach the estimated strain of the candidate reported by~\cite{ming2025x}.\\

\emph{SHV results:}
All candidates surviving the veto steps of the SHV pipeline are consistent with noise fluctuations, and no evidence of CWs is found in the LIGO O4a dataset. 
None of them survived the follow-up procedures described in Appendix \ref{appendix:SHV_followup}.
Constraints on the GW strain amplitude are derived using a scaling relation described in Appendix~\ref{appendix:SHV_UL}, supported by numerical upper limit calculations based on software injections. The results are shown in Appendix~\ref{appendix:SHV_UL}.
Across all analyzed targets, we state here the best upper limit achieved at 95\% confidence level  $h_0^{95\%} = 6.14 \times 10^{-26}$ at $f= 164$ Hz from G65.7+1.2.
The best depth at 200 Hz across all sources is 51.7 $\text{Hz}^{-1/2}$ (G65.7+1.2) while Vela Jr. reaches a depth of 28.2 at the same frequency. The depth reached by SVH is not enough to probe the candidate in~\cite{ming2025x}. Furthermore, the search setup used for G347.3$-$0.5 is not able to produce candidates with first order spin-down parameters compatible with the candidate.\\

\emph{DHV results:}
Without finding evidence of CWs in the dual-harmonic Viterbi search (see Appendix~\ref{appendix:dhv_final_veto}), we take a frequentist approach and estimate the 95\% confidence strain sensitivity, denoted by $h_0^{95\%}$, in a frequency sub-band, and extend this estimate to the whole frequency band using an analytical scaling as described in Appendix~\ref{appendix:DHV_UL}.
Note that the strain sensitivity quoted in this pipeline is based on a different signal model from other pipelines [see Eqs.~(\ref{eqn:h2p})--(\ref{eqn:h1c})].
In this signal model, the strain sensitivity depends on the star's wobble angle and orientation. 
Thus, we consider a specific scenario, representing the worst-case for single-band searches, where tracking the two frequency bands simultaneously offers the most significant sensitivity improvement compared to a single-band search. This scenario assumes $\theta = 45$~deg and $\cos \iota = 0$, corresponding to linearly polarized signals at both $\fstar$ and $2\fstar$~\citep{Sun2019}.
In Appendix~\ref{appendix:DHV_UL} is reported the estimated $h_0^{95\%}$ in all sub-bands.
The best $h_0^{95\%}$ values over the full frequency band searched are $1.82 \times 10^{-25}$ at $\fstar = 213.25$~Hz and $2.12 \times 10^{-25}$ at $\fstar = 144.25$~Hz, for $\Tcoh=12$~hr and 9~hr, respectively.
These results are obtained from randomized sky positions and hence apply to all sources using the same $\Tcoh$.\footnote{The sensitivity is dominated by the length of $\Tcoh$ (for a fixed $\Tobs$) rather than the sky position of the source in the dual-harmonic Viterbi pipeline~\citep{Abbott2021_O3aSNR}.}
The sensitivity depths for this pipeline are 13.9~Hz$^{-1/2}$ for $\Tcoh=12$~hr and 12.4~Hz$^{-1/2}$ for $\Tcoh=9$~hr.\\

\emph{Weave results:}
Using Weave, we find no significant evidence for a CW signal from the Central Compact Objects in the SNRs \casa, \vela, and \targetg (see Appendix~\ref{appendix: weave follow-up}).
Given the absence of a detection, we determine 95\%\ confidence level frequentist upper limits on strain amplitude $h_0^{95\%}$ for each 1-Hz band,
excluding saturated 0.1-Hz sub-bands.
The upper limit results for all the targets are  reported in Appendix~\ref{appendix: weave ul injections} including the detailed procedure to compute them.
The resulting frequency-dependent sensitivity depths have values at 200 Hz of $\depthc=87.0~\text{Hz}^{-1/2}$, $\depthv=93.2~\text{Hz}^{-1/2}$, and $\depthg=93.1~\text{Hz}^{-1/2}$. In comparison with~\cite{ming2025x}, although our results are more sensitive at higher frequencies, our sensitivity at the candidate frequency reported by \cite{ming2025x} is lower. Consequently, our search does not reach the quoted strain estimated for this candidate, which is consistent with its non-detection.\\

For all the results we observe an improvement with respect to O3a searches, compatible with the longer observation period in O4a (8 months compared to 6 months in O3a) and an improved detector sensitivity (of at least a factor of 1.5). Further improvements are observed for SHV, where longer coherence times and an adapted search setup have been applied here. For Weave, an additional improvement with respect to O3a is due to the lower detection statistic threshold used to define initial candidates for follow-up. 

\subsection{Astrophysical constraints}
\label{sec:astrolimits}
\begin{figure*}[htb!]
	\centering
	\subfigure[][Equatorial ellipticity]
	{
		\label{fig:BSD_ellipticity}
		\scalebox{0.34}{\includegraphics{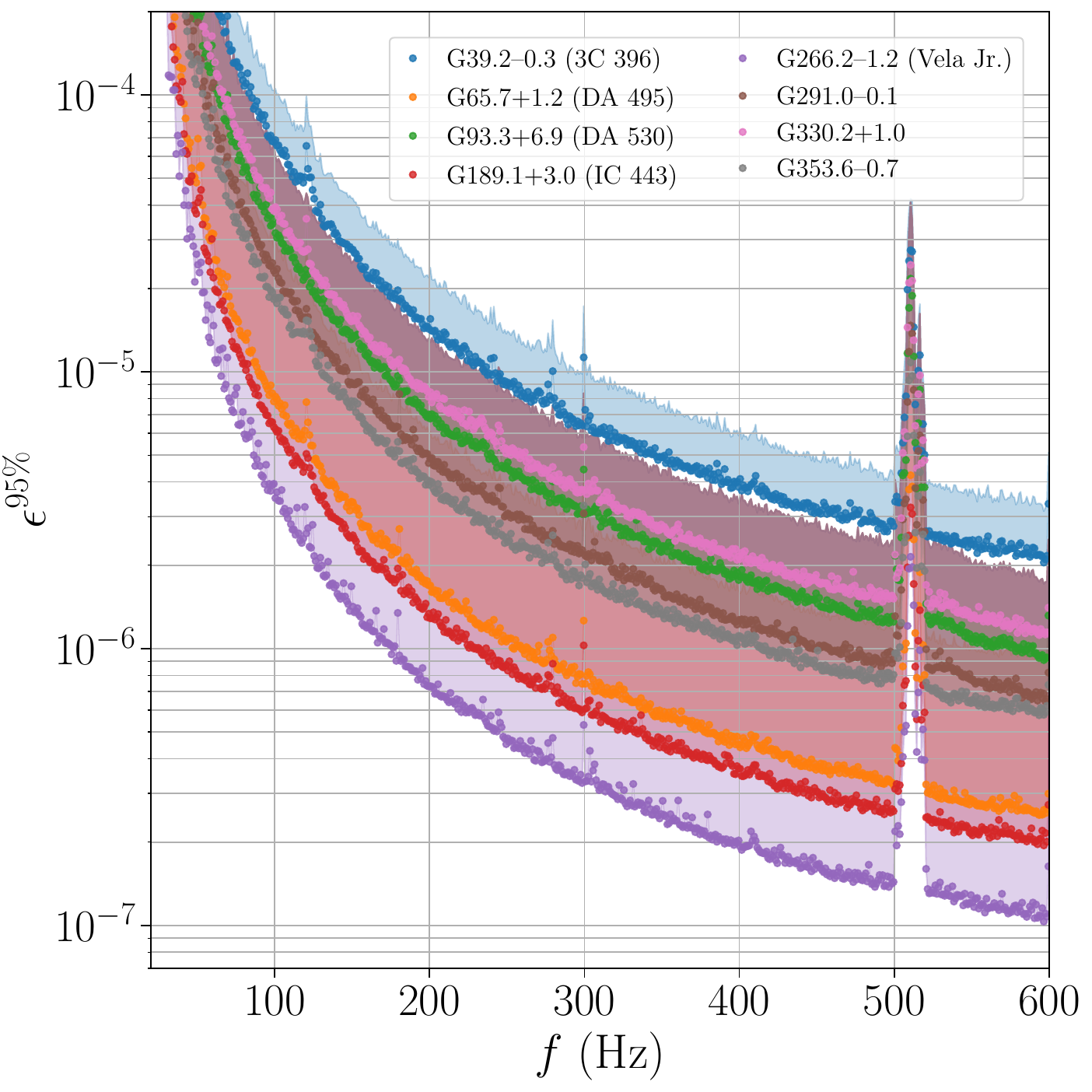}}
	}
	\subfigure[][$r$-mode amplitude $\alpha$]
	{
		\label{fig:BSD_rmodes}
		\scalebox{0.34}
        {\includegraphics{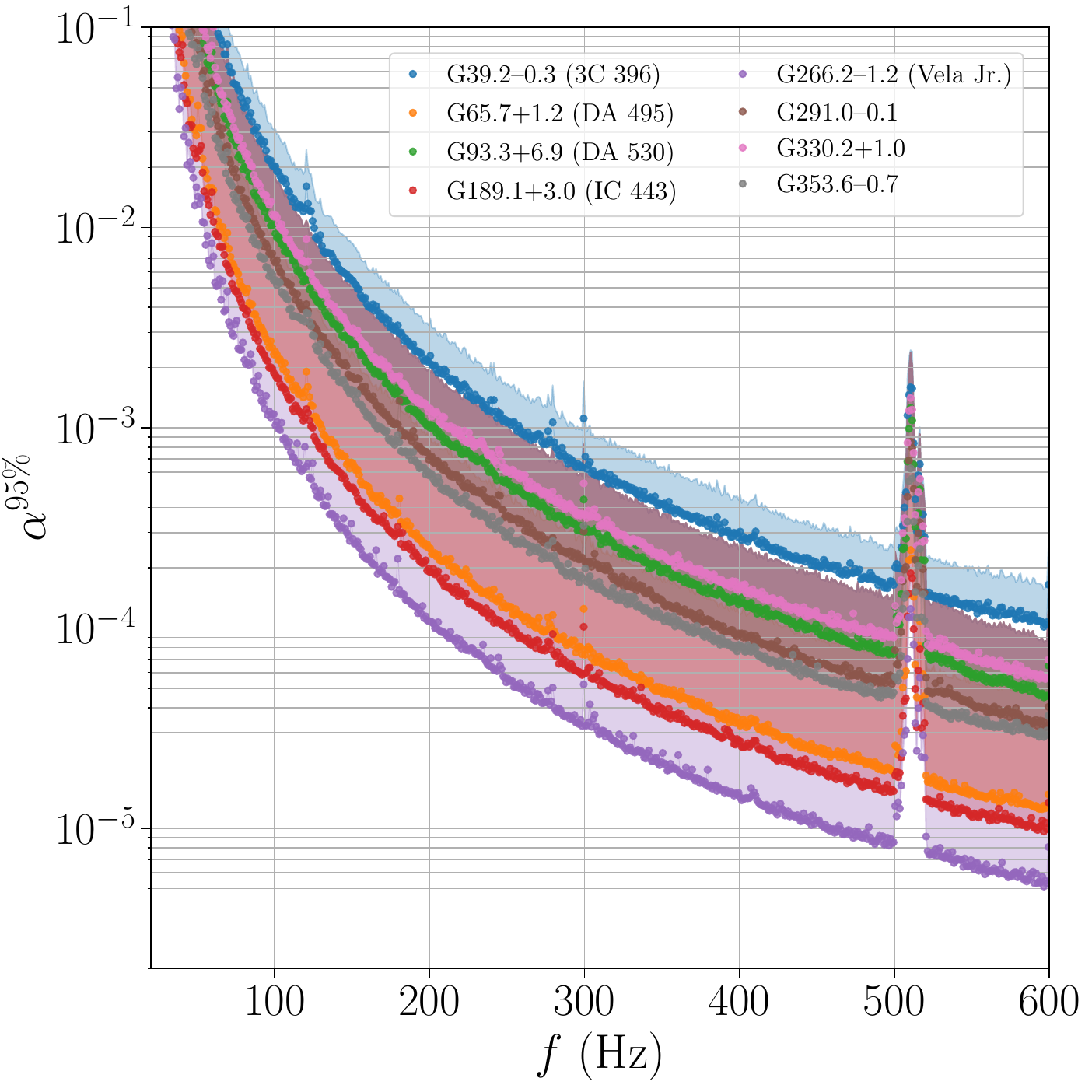}}
	}
	\caption{Equatorial ellipticity (left) and $r$-mode amplitude $\alpha$ (right) with 95\% confidence for all the BSD targets. Filled circles show the ellipticity or $r$-mode amplitude computed using the minimum estimated source distance $D_{\min}$. Shaded regions indicate the full range of values obtained when varying the source distance from $D_{\min}$ to $D_{\max}$ (see Table~\ref{tab:targets}). The limits are derived from the upper limits curves in Fig.~\ref{fig:BSD_ULall}.}	
\label{fig:BSDastro}
\end{figure*}

\begin{figure*}[htb!]
	\centering
	\subfigure[][Equatorial ellipticity]
	{
		\scalebox{0.34}{\includegraphics{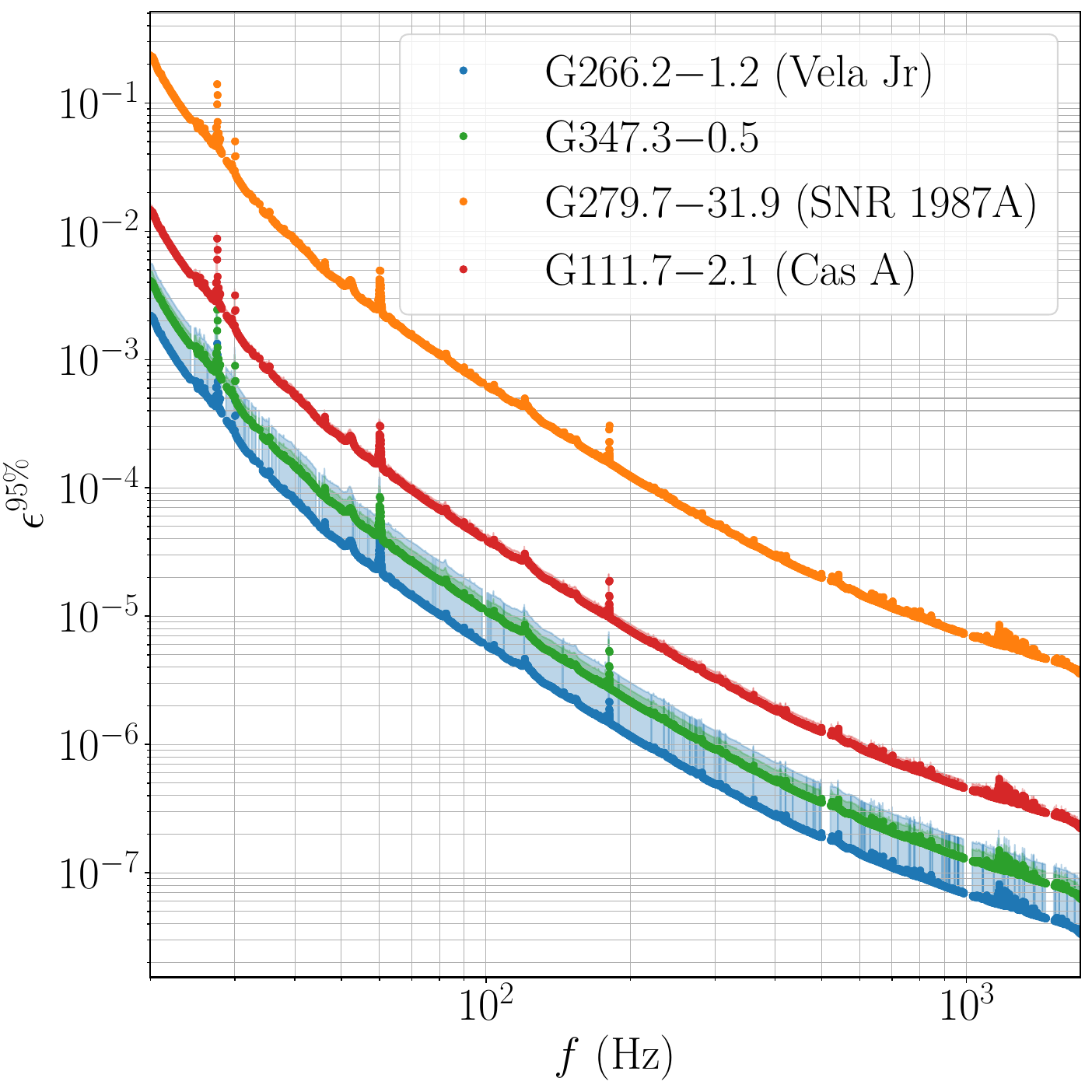}}
	}
	\subfigure[][$r$-mode amplitude $\alpha$]
	{
		\scalebox{0.34}
        {\includegraphics{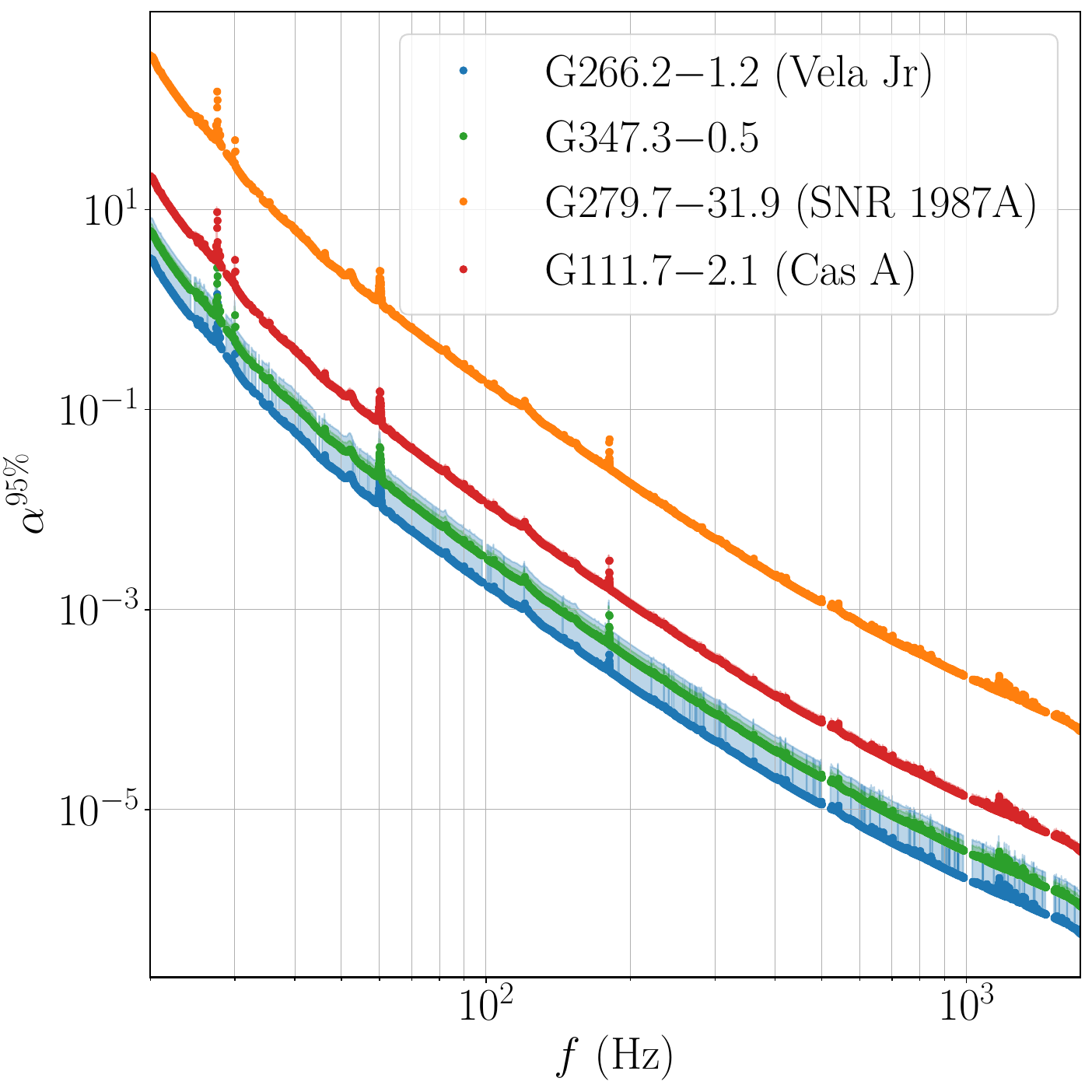}}
	}
	\caption{Equatorial ellipticity (left) and $r$-mode amplitude $\alpha$ (right) with 95\% confidence for all targets studied in the PyStoch search. The filled circles indicate the minimum $\epsilon^{95\%}$ for each SNR, and the shaded regions span the full range of ellipticity values obtained by varying the distance between $D_{\min}$ and $D_{\max}$ (see Table~\ref{tab:targets}).}	
\label{fig:PyStochastro}
\end{figure*}

\begin{figure*}[htb!]
	\centering
	\subfigure[][Equatorial ellipticity]
	{
		\scalebox{0.34}{\includegraphics{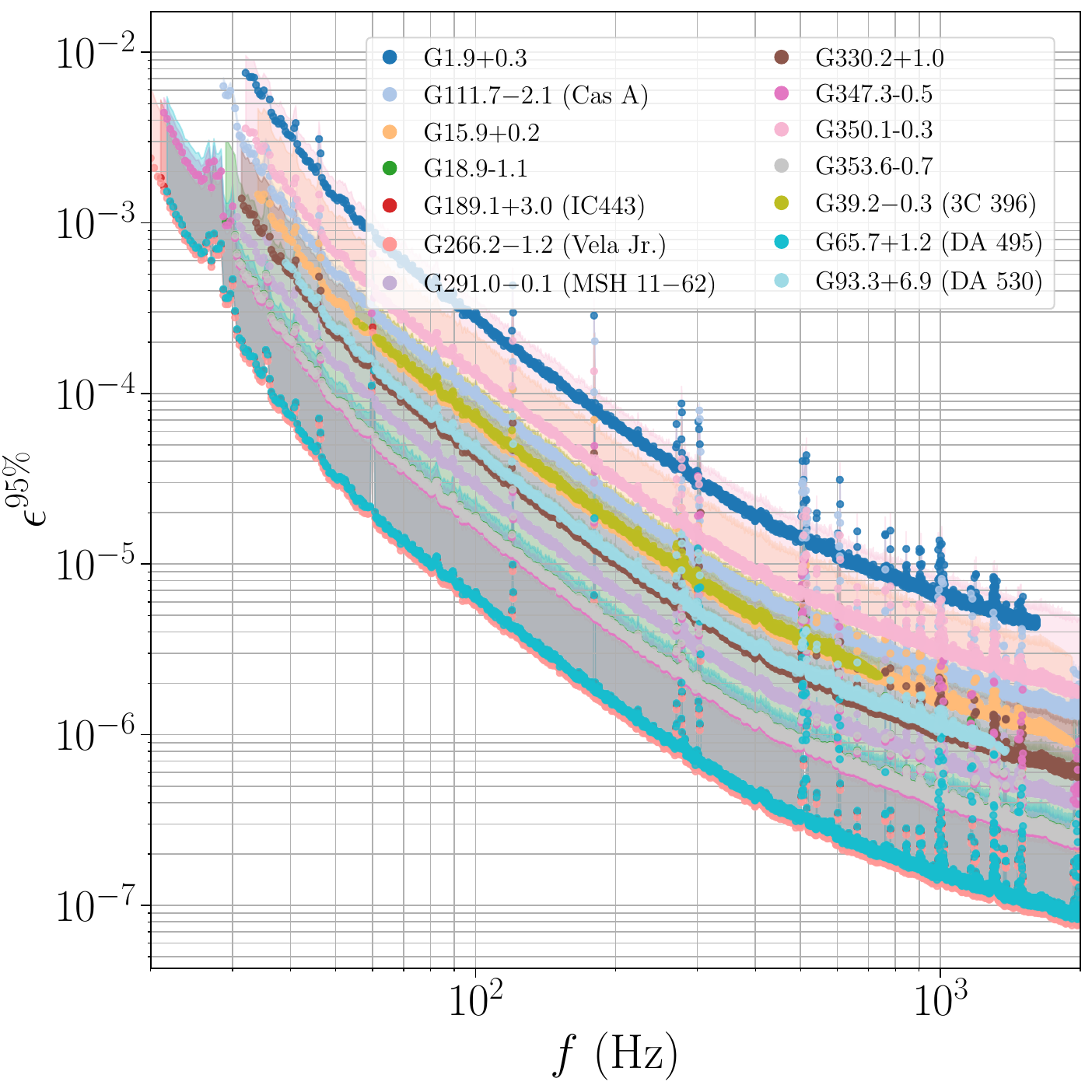}}
	}
	\subfigure[][$r$-mode amplitude $\alpha$]
	{
		\scalebox{0.34}{\includegraphics{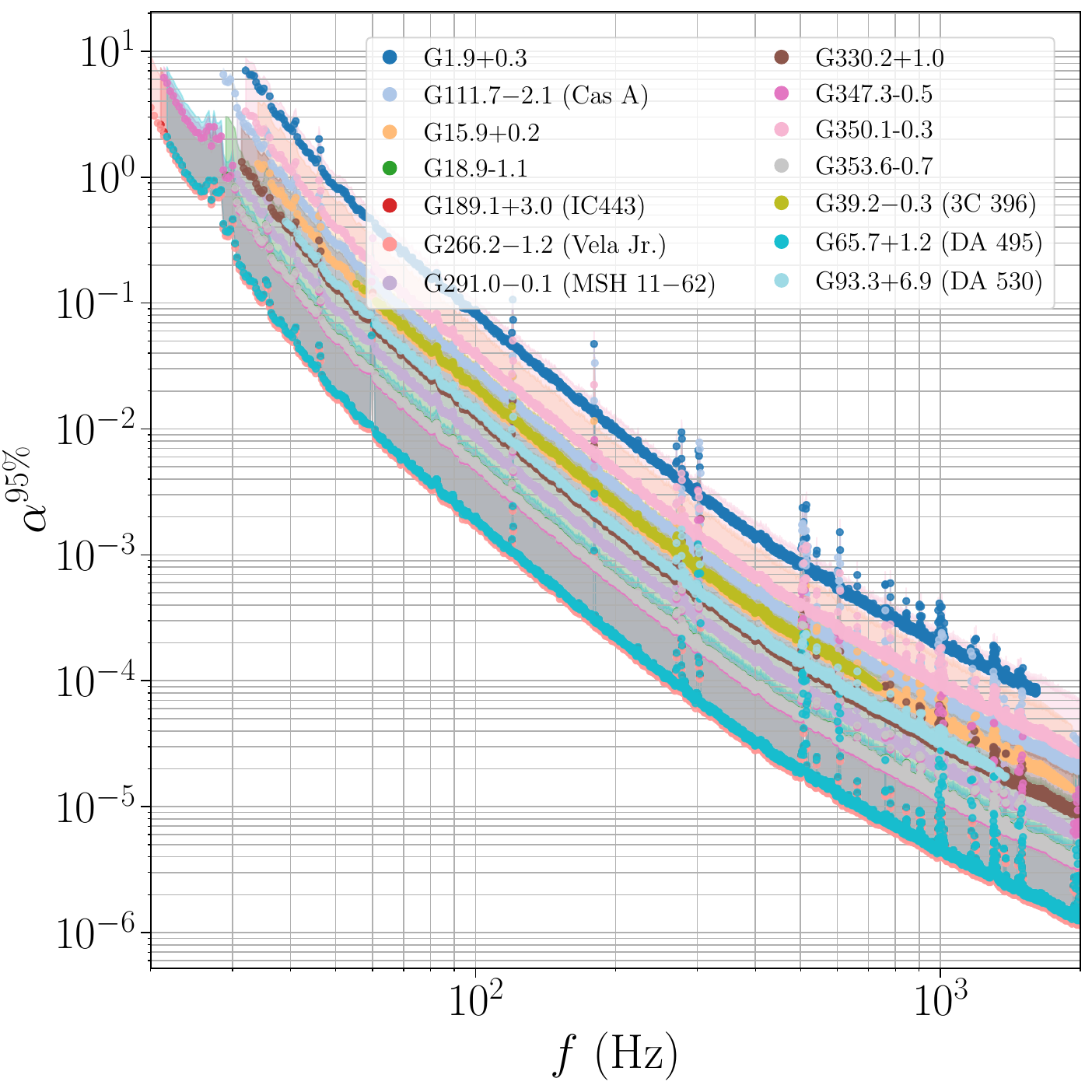}}
	}
	\caption{Equatorial ellipticity (left) and $r$-mode amplitude $\alpha$ (right) with 95\% confidence for all the SHV targets. The filled circles indicate the minimum $\epsilon^{95\%}$ and $\alpha^{95\%}$ for each SNR, and the shaded regions span the full range of ellipticity and \rmode amplitude values obtained by varying the distance between $D_{\min}$ and $D_{\max}$ (see Table~\ref{tab:targets}).}	
\label{fig:SHVastro}
\end{figure*}

\begin{figure*}[htb!]
	\centering
	\subfigure[][Equatorial ellipticity]
	{
		\label{fig:weave_ellipticity}
		\scalebox{0.34}{\includegraphics{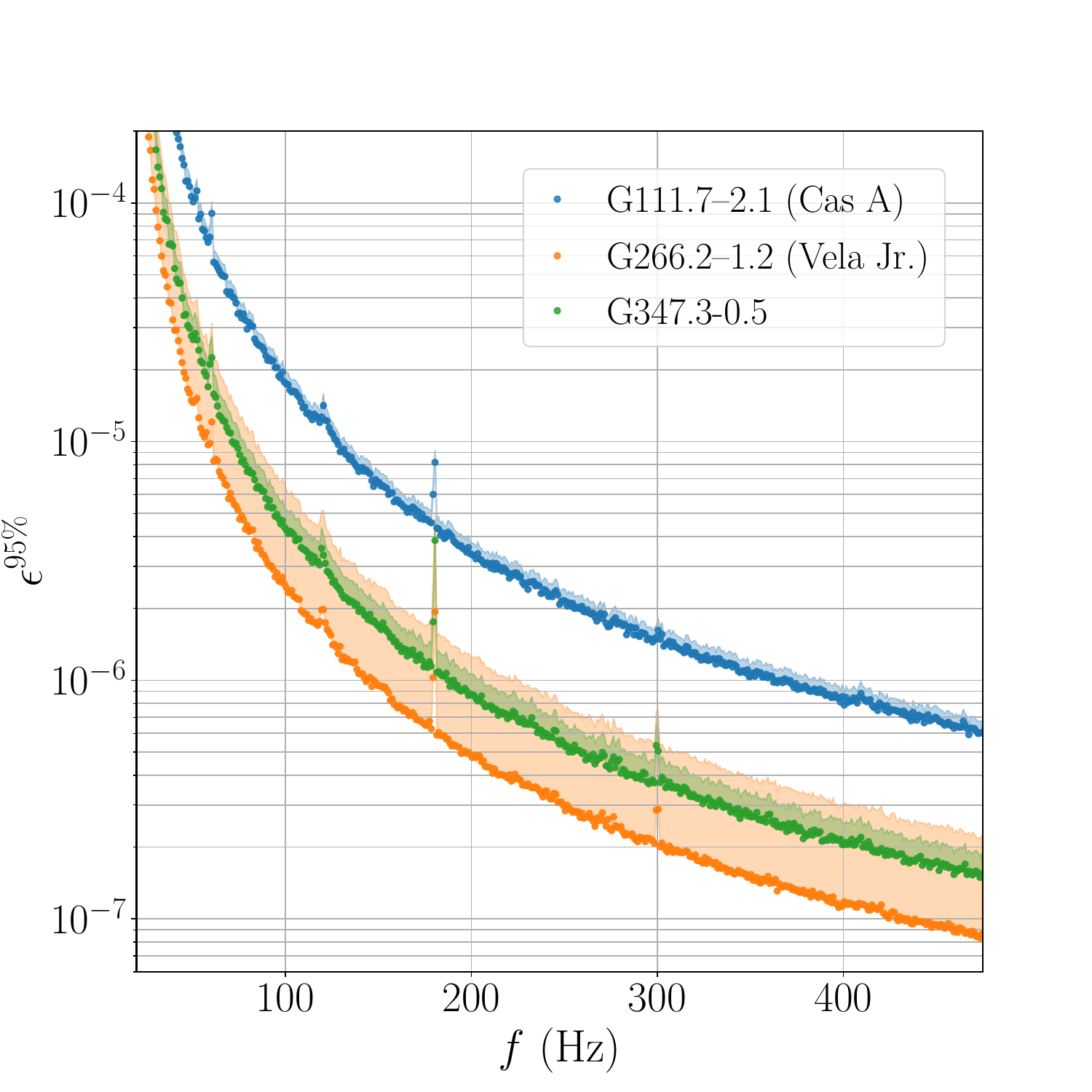}}
	}
	\subfigure[][$r$-mode amplitude $\alpha$]
	{
		\label{fig:weave_rmodes}
		\scalebox{0.34}{\includegraphics{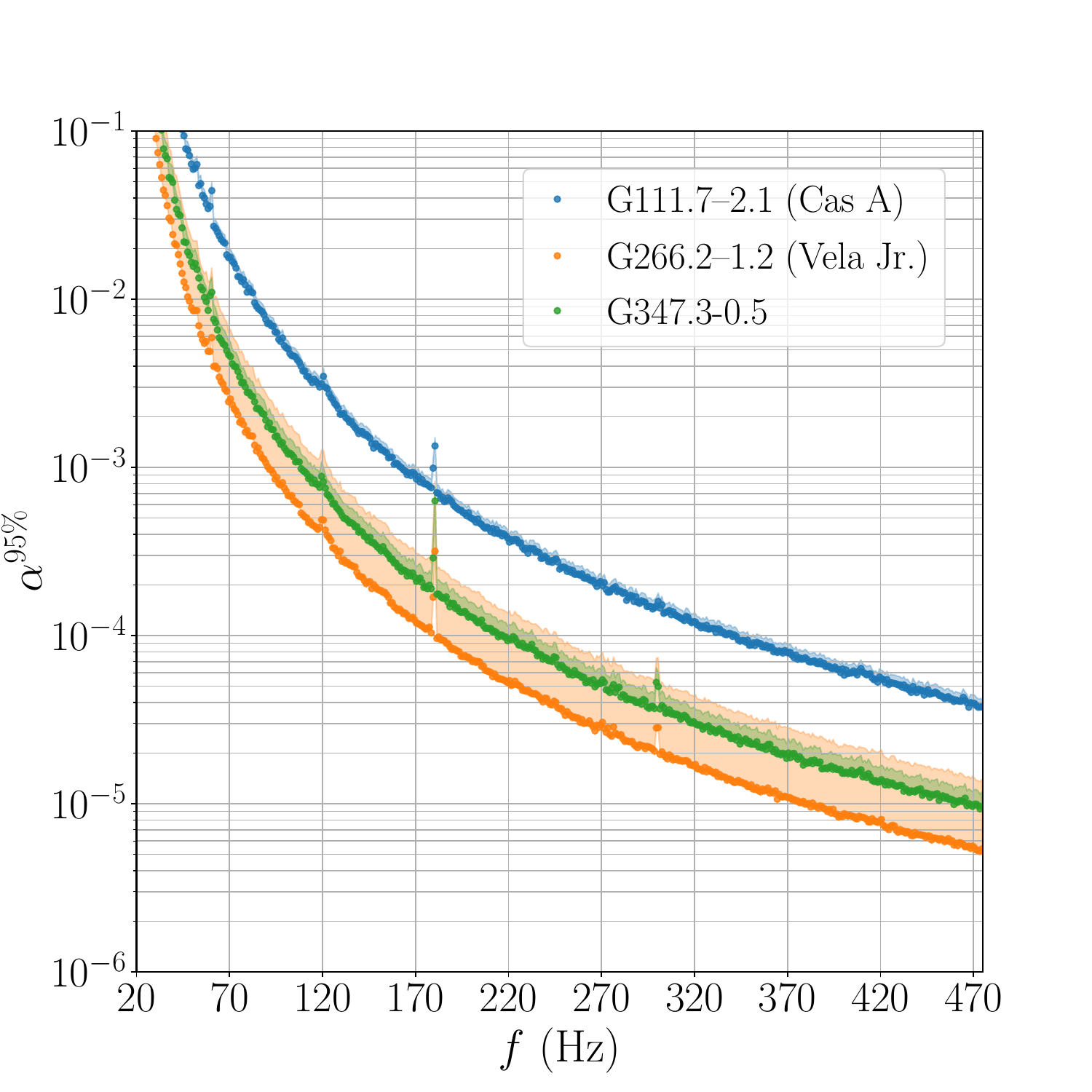}}
	}
	\caption{Equatorial ellipticity and $r$-mode amplitude $\alpha$ upper limits (95\%\ confidence level) from the Weave searches in each 1-Hz band for \casa (blue), \vela (orange), and \targetg (green), derived from the strain amplitude upper limits shown in Fig.~\ref{fig:weave_sensitivities}. 
    The shaded region indicates the resulting ellipticity/$r$-mode amplitude range calculated from the full estimated distance range for each SNR (see Table~\ref{tab:targets}), and the solid dots correspond to the minimum distance estimate.}	
\label{fig:weave_astro}
\end{figure*}

\begin{figure*}[htb!]
	\centering
	\subfigure[]
	{
		\label{fig:eps-12hr}
		\scalebox{0.34}{\includegraphics{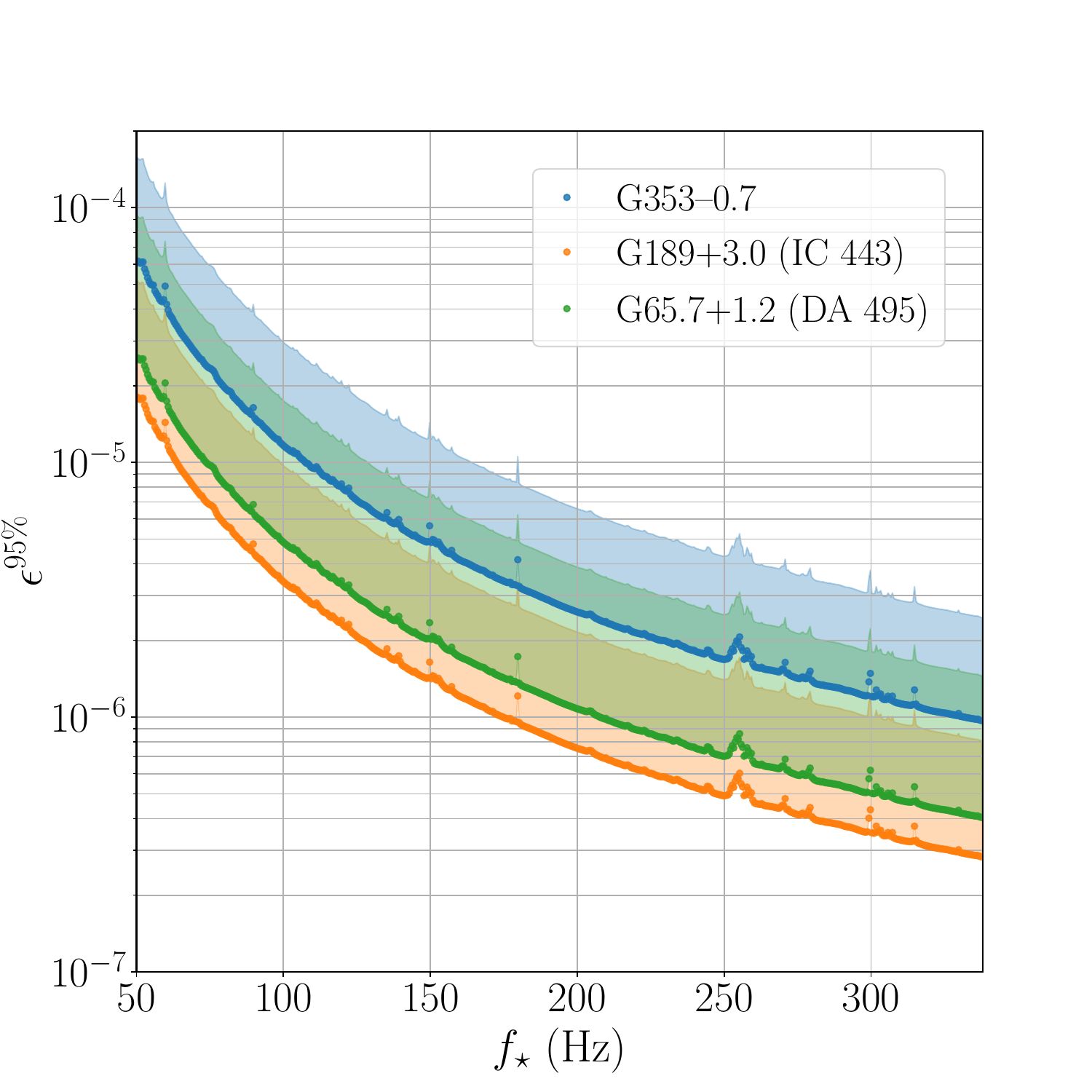}}
	}
	\subfigure[]
	{
		\label{fig:eps-9hr}
		\scalebox{0.34}{\includegraphics{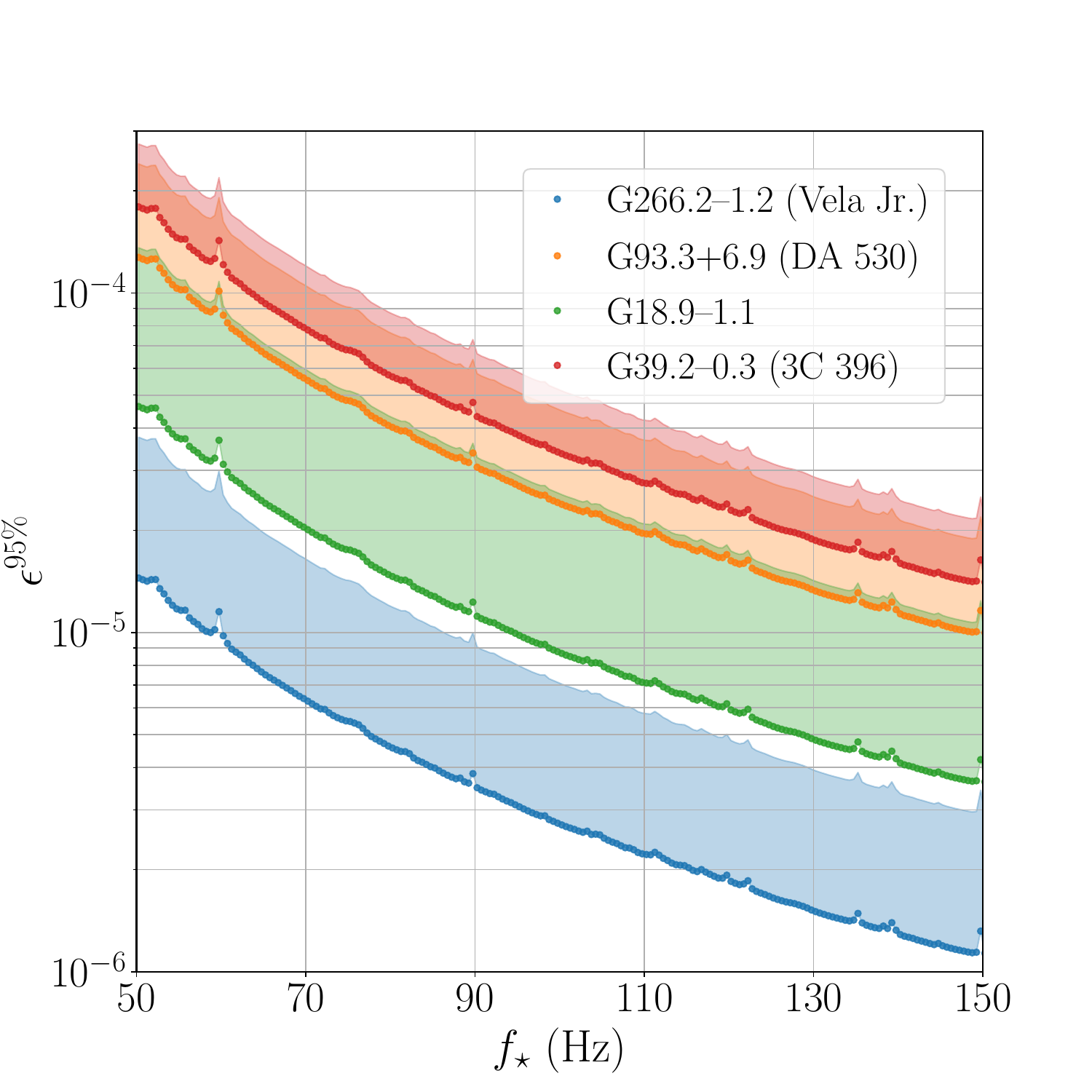}}
	}
	\caption{Constraints on the NS ellipticity with 95\% confidence from the dual-harmonic Viterbi search as a function of $\fstar$ for (a) $\Tcoh=12$~hr and (b) $\Tcoh=9$~hr, converted from the $h_0^{95\%}$ values in Figure~\ref{fig:viterbi1f2fUL}. The shaded region indicates the resulting ellipticity range calculated from the full estimated distance range for each SNR (see Table~\ref{tab:targets}), and the solid dots correspond to the minimum distance estimate.}
	\label{fig:viterbi1f2feps}
\end{figure*}

The GW strain amplitude can be converted into the fiducial ellipticity of the NS, $\epsilon$~\citep{Jaranowski1998}, and the \rmode amplitude parameter $\alpha$ (if an \rmode emission is assumed~\citealt{Owen2010}).
Given $h_0^{95\%}$, we constrain the ellipticity of the NS in terms of the GW frequency $f$ via
\begin{equation}
  \label{eq:ellipticity}
  \epsilon  = 9.46\times10^{-5}\Bigl(\frac{h_0}{10^{-24}}\Bigr)\Bigl(\frac{D}{1\>{\rm kpc}}\Bigr)
\Bigl(\frac{100\>{\rm Hz}}{f}\Bigr)^2,
\end{equation}
assuming that the moment of inertia with respect to the rotation axis ($I_{zz}$ for a perpendicular biaxial rotor) is $10^{38}$ kg m$^2$. Here we have considered the standard scenario of a single harmonic emission at $2\fstar$, valid for all the pipelines (Figures~\ref{fig:BSDastro}--\ref{fig:weave_astro}) except for DHV.  For the specific dual-harmonic scenario, the GW emission is at both $\fstar$ and $2\fstar$, hence we can write Eq.~(\ref{eq:ellipticity}) in terms of the spin frequency of the star $\fstar$,
\begin{equation} \label{eqn:eps_1f2f} \epsilon = 2.36\times 10^{-5} \left(\frac{h_0}{10^{-24}}\right) \left(\frac{D}{1{\rm~kpc}}\right) \left(\frac{100{\rm~Hz}}{\fstar}\right)^2.
\end{equation}
Figure~\ref{fig:viterbi1f2feps} shows the constraints on $\epsilon$ as a function of $\fstar$ obtained for the DHV search.
These constraints are converted from $h_0^{95\%}$ and are obtained for a specific scenario with source properties $\theta=45$~deg and $\cos\iota = 0$. Therefore, the limits on $\epsilon$ in Figure~\ref{fig:viterbi1f2feps} are not directly comparable to the results obtained in other conventional searches where $\theta=90$~deg and the emission is only at $2\fstar$ and reported in Figures~\ref{fig:BSDastro}--\ref{fig:weave_astro}.

We can also convert $h_0^{95\%}$ to limits on the amplitude of \rmode oscillations via~\citep{Owen2010}:
\begin{equation}
  \label{eq:rmodes}
  \alpha  \simeq 0.028\Bigl(\frac{h_0}{10^{-24}}\Bigr)\Bigl(\frac{D}{1\>{\rm kpc}}\Bigr)
\Bigl(\frac{100\>{\rm Hz}}{f}\Bigr)^3.
\end{equation}

Figures~\ref{fig:BSDastro}–\ref{fig:weave_astro} show the constraints derived from the BSD, PyStoch, SHV, and Weave 95\% confidence-level upper limits for the ellipticity (left) and the \rmode amplitude (right). Solid curves correspond to $D = D_{\rm min}$, while shaded regions indicate the constraints obtained for distances in the range $D_{\rm min} \leq D \leq D_{\rm max}$. Distances are taken from Table~\ref{tab:targets}. 

We do not infer $r$-mode amplitudes for the DHV pipeline since the signal model adopted in the dual-harmonic search cannot be interpreted as current quadrupole emission from an $r$-mode.

The strongest ellipticity and $r$-mode constraints are obtained at the highest frequencies. Indeed, for all sources and targets we are able to reach values below $\epsilon < 10^{-5}$ at the high end of the frequency band. Despite uncertainties in the source distances, we are approaching or exceeding the rough theoretical upper limit for normal NSs~\citep{Johnson-McDaniel2013eps}. The closest source, G266.2$-$1.2 (Vela Jr.), dramatically surpasses this limit, with ellipticity constraints reaching below $\epsilon \approx 10^{-7}$ across all pipelines.

These constraints are, however, model dependent: uncertainties in the stellar geometry and composition, including the internal equation of state, the neutron-star mass, the moment of inertia, and the magnetic field strength, can significantly affect the inferred values introducing uncertainties of up to a factor of approximately two to three. Despite these limitations, the improved detector sensitivities allow us to place tighter limits than previous O3 searches over a broader parameter space. In particular, for SNR~1987A, the constraints improve upon the O3 modeled search results in~\citep{Owen_2024} by a factor of $\sim 1.5$, consistent with the different integration time and the improved detector sensitivity used in this search.
The most stringent constraints on the $r$-mode amplitude obtained above $\sim 200$~Hz strongly surpass the theoretical prediction level of $\alpha \sim 10^{-3}$ expected from nonlinear saturation mechanisms~\citep{Bondarescu2009rmode}, reaching values of $\alpha \sim 10^{-6}$ at the highest frequencies covered by the SHV pipeline.

\section{Conclusions}
\label{sec:conclusions}
In this work, we have presented the results of a comprehensive multi-pipeline search for CWs in SNRs potentially hosting young NSs, using data from the first eight months of the O4 LIGO-Virgo-KAGRA run. Our analysis employed complementary pipelines — BSD, PyStoch, SHV, DHV, and Weave — each targeting distinct signal models, parameter space and offering varied sensitivity trade-offs. No evidence of a CW signal is found; all candidates are consistent with noise or eliminated by follow-up analyses. 

We place new upper limits on the GW strain amplitude \( h_0 \) at 95\% confidence level beating previous O3 results. The BSD pipeline sets limits across 1-Hz sub-bands from 10 to 600~Hz, surpassing indirect energy conservation limits in several sources, notably for G65.7+1.2, G189.1+3.0, and Vela Jr.; this limit is beaten by a factor of $\sim 20, 25, 50$, respectively, assuming the most optimistic scenario of a closeby and young source. However, these limits are surpassed even in the worst case configuration for 6 out of 8 sources in the BSD search. PyStoch results, spanning 20--1726~Hz at varying frequency resolutions, present the results from the search for CW emission from the SNR~1987A using an unmodeled approach, updating the limits found in the O3 search~\citep{Nostro}.
The Viterbi pipelines provide constraints that incorporate spin wandering effects (single-harmonic) and simultaneous emission at \( f_\star \) and \( 2 f_\star \) (dual-harmonic). 
Single-harmonic upper limits reach $h_0^{95\%} = 6.14 \times 10^{-26}$ at $f= 164.00$~Hz from G65.7+1.2.
Dual-harmonic upper limits reach \( h_0^{95\%} = 1.82 \times 10^{-25} \) at \( f_\star = 213.25 \)~Hz (12-hr coherence) and \( 2.12 \times 10^{-25} \) at \( f_\star = 144.25 \)~Hz (9-hr coherence), consistent with analytic expectations.
The Weave pipeline’s high-resolution follow-up confirms all initial candidates are consistent with noise or instrumental artifacts reaching the best $h_0^{95\%}$ limit on the signal strain \(\approx 4.33 \times 10^{-26}\) near 283~Hz for \casa, \(\approx 3.92 \times 10^{-26}\) near 266~Hz for \vela, and \(\approx 3.85 \times 10^{-26}\) near 378~Hz for \targetg.

Together, these results represent a significant advance in search sensitivity compared with previous searches for CWs from young NSs in nearby SNRs. 
These results provide the strongest upper limits to date across the full analyzed frequency band, even relative to computationally intensive searches in~\cite{Ming_2022,Morales_2025}.
While no signals are detected, the improved upper limits place stronger constraints on NS deformation and energy loss models. Future observing runs with longer data sets and enhanced detector sensitivity, combined with multi-pipeline strategies demonstrated here, will further improve detection prospects.

\section{Acknowledgments}
%
This material is based upon work supported by NSF's LIGO Laboratory, which is a
major facility fully funded by the National Science Foundation.
The authors also gratefully acknowledge the support of
the Science and Technology Facilities Council (STFC) of the
United Kingdom, the Max-Planck-Society (MPS), and the State of
Niedersachsen/Germany for support of the construction of Advanced LIGO 
and construction and operation of the GEO\,600 detector. 
Additional support for Advanced LIGO was provided by the Australian Research Council.
The authors gratefully acknowledge the Italian Istituto Nazionale di Fisica Nucleare (INFN),  
the French Centre National de la Recherche Scientifique (CNRS) and
the Netherlands Organization for Scientific Research (NWO)
for the construction and operation of the Virgo detector
and the creation and support  of the EGO consortium. 
The authors also gratefully acknowledge research support from these agencies as well as by 
the Council of Scientific and Industrial Research of India, 
the Department of Science and Technology, India,
the Science \& Engineering Research Board (SERB), India,
the Ministry of Human Resource Development, India,
the Spanish Agencia Estatal de Investigaci\'on (AEI),
the Spanish Ministerio de Ciencia, Innovaci\'on y Universidades,
the European Union NextGenerationEU/PRTR (PRTR-C17.I1),
the ICSC - CentroNazionale di Ricerca in High Performance Computing, Big Data
and Quantum Computing, funded by the European Union NextGenerationEU,
the Comunitat Auton\`oma de les Illes Balears through the Conselleria d'Educaci\'o i Universitats,
the Conselleria d'Innovaci\'o, Universitats, Ci\`encia i Societat Digital de la Generalitat Valenciana and
the CERCA Programme Generalitat de Catalunya, Spain,
the Polish National Agency for Academic Exchange,
the National Science Centre of Poland and the European Union - European Regional
Development Fund;
the Foundation for Polish Science (FNP),
the Polish Ministry of Science and Higher Education,
the Swiss National Science Foundation (SNSF),
the Russian Science Foundation,
the European Commission,
the European Social Funds (ESF),
the European Regional Development Funds (ERDF),
the Royal Society, 
the Scottish Funding Council, 
the Scottish Universities Physics Alliance, 
the Hungarian Scientific Research Fund (OTKA),
the French Lyon Institute of Origins (LIO),
the Belgian Fonds de la Recherche Scientifique (FRS-FNRS), 
Actions de Recherche Concert\'ees (ARC) and
Fonds Wetenschappelijk Onderzoek - Vlaanderen (FWO), Belgium,
the Paris \^{I}le-de-France Region, 
the National Research, Development and Innovation Office of Hungary (NKFIH), 
the National Research Foundation of Korea,
the Natural Sciences and Engineering Research Council of Canada (NSERC),
the Canadian Foundation for Innovation (CFI),
the Brazilian Ministry of Science, Technology, and Innovations,
the International Center for Theoretical Physics South American Institute for Fundamental Research (ICTP-SAIFR), 
the Research Grants Council of Hong Kong,
the National Natural Science Foundation of China (NSFC),
the Israel Science Foundation (ISF),
the US-Israel Binational Science Fund (BSF),
the Leverhulme Trust, 
the Research Corporation,
the National Science and Technology Council (NSTC), Taiwan,
the United States Department of Energy,
and
the Kavli Foundation.
The authors gratefully acknowledge the support of the NSF, STFC, INFN and CNRS for provision of computational resources.

This work was supported by MEXT,
the JSPS Leading-edge Research Infrastructure Program,
JSPS Grant-in-Aid for Specially Promoted Research 26000005,
JSPS Grant-in-Aid for Scientific Research on Innovative Areas 2402: 24103006,
24103005, and 2905: JP17H06358, JP17H06361 and JP17H06364,
JSPS Core-to-Core Program A.\ Advanced Research Networks,
JSPS Grants-in-Aid for Scientific Research (S) 17H06133 and 20H05639,
JSPS Grant-in-Aid for Transformative Research Areas (A) 20A203: JP20H05854,
the joint research program of the Institute for Cosmic Ray Research,
University of Tokyo,
the National Research Foundation (NRF),
the Computing Infrastructure Project of the Global Science experimental Data hub
Center (GSDC) at KISTI,
the Korea Astronomy and Space Science Institute (KASI),
the Ministry of Science and ICT (MSIT) in Korea,
Academia Sinica (AS),
the AS Grid Center (ASGC) and the National Science and Technology Council (NSTC)
in Taiwan under grants including the Science Vanguard Research Program,
the Advanced Technology Center (ATC) of NAOJ,
and the Mechanical Engineering Center of KEK.

Additional acknowledgements for support of individual authors may be found in the following document: \\
\texttt{https://dcc.ligo.org/LIGO-M2300033/public}.
For the purpose of open access, the authors have applied a Creative Commons Attribution (CC BY)
license to any Author Accepted Manuscript version arising.
We request that citations to this article use 'A. G. Abac {\it et al.} (LIGO-Virgo-KAGRA Collaboration), ...' or similar phrasing, depending on journal convention. This article has been assigned document number LIGO-P2500356.



\appendix

\section{Details on the BSD search}
\label{appendix:BSDmethod}

\subsection{Search implementation}
\label{appendix:BSDserchsetuptargets}
The input data of the BSD semi-coherent search consists of a complex-valued sub-sampled time series, the BSD files~\citep{BSD}.
For each target, the detector's time series is partially corrected for the Doppler effect, assuming the known sky location of the source. This correction is repeated for each 1-Hz sub-band using its central frequency (as detailed in \citealt{DirectedBSD}).
From these partially corrected time series, we construct time-frequency peakmaps by extracting local maxima above a given threshold in the equalized spectrum~\citep{FFTpeakmaps}. The peakmaps are then fed into the FH transform, which maps the time-frequency peaks onto the source's frequency and spin-down plane $(f_0,\dot{f_0})$ at a given reference time $t_0$. The second-order spin-down parameter $\ddot{f_0}$ is always assumed equal to zero.
The first order spin-down range is defined as $-f_{\rm max}/\tage \leq \dot{f} \leq 0.1 f_{\rm max}/\tage$, with $f_{\rm max}$ the upper frequency of each 10-Hz band. Given that the second-order spin-down is neglected, and we fix a maximum frequency of $600~\rm Hz$ for all the targets, the minimum age of the source will be $3.8~\rm kyr$ (see e.g., Eq. (B2) of~\citealt{Abbott2021_O3aSNR}).

Each FH map is built with a  resolution equal to the template grid spacing as  \citep{Abbott2021_O3aSNR}:
\begin{eqnarray}
\delta f_{\rm FH} = \frac{1}{T_{\rm coh}K_{f}},
\,\,\delta \dot{f}_{\rm FH}=\frac{1}{T_{\rm coh}T_{\rm obs}K_{\dot{f}}}, 
\label{eq:deltadf_O4} 
\end{eqnarray} 
where $T_{\rm coh}$ is the coherence time, $T_{\rm obs}$ the observational time span and $K_{f}=10$ and $K_{\dot{f}}=2$ are the over-resolution factors following \cite{FrequencyHoughmethod}. Since the coherence time scales with the maximum frequency as $T_{\rm coh} \propto 1/\sqrt{f_{\rm max}}$, the grid resolution varies across frequency bands. 
The coherence time used varies from 17.8 hours at the lowest frequencies to 3.3 hours at 600 Hz. 
We search both negative and positive spin-down to allow for possible spin-up scenarios. 
The list of targets and search configurations is listed in Table~\ref{tab:parabsd}.

\begin{table*}[!tbh]
	\centering
	\setlength{\tabcolsep}{3pt}
	\renewcommand\arraystretch{1.2}
	\begin{tabular}{lcccc}
		\toprule
		Source$^a$ &  Minimum $\tage$ (kyr) & $T_{\rm coh}$ (hr)  & $f$ (Hz) & $\dot{f}$ $(\rm{Hz\,s^{-1}})$ \\
		       &                              & (@100 Hz)           &          &   (@100 Hz)\\
		\midrule
		G65.7+1.2, G189.1+3.0, G266.2$-$1.2,  & 3.8   & 8 & [10, 600]  & $[-9.18 \times 10^{-10}, 9.18\times 10^{-11}]$  \\    
        G353.6$-$0.7, G93.3+6.9, & & & & \\
        G291.0$-$0.1, G330.2+1.0 & & & & \\
\hline
		G39.2$-$0.3 						   & 4.7 & 8 & [10, 600] & $[-7.42\times 10^{-10}, 7.42\times 10^{-11}]$  \\
		\bottomrule
	\end{tabular}
	\caption{Sources searched in the BSD analysis (Section~\ref{sec:BSD}) and the parameter space covered. The coherence time and the spin-down/up range scale with the maximum frequency in each 10-Hz frequency band. For each source, we report the coherence time, $T_{\rm coh}$, and spin-down/up range for $\dot{f}$, used for the frequency band [90, 100] Hz where $f_{\rm max}=100$ Hz. \\
    $^a$ G291.0$-$0.1 and G330.2+1.0 are two newly added sources in the BSD O4a search, with respect to the BSD O3a search.}
	\label{tab:parabsd}
\end{table*}

As in previous searches, candidates with the highest FH number count are selected from a final map built by summing all monthly FH maps for a given source and frequency range. Candidates are selected independently from each detector, and coincident candidates are then passed to the postprocessing and follow-up steps. 
We additionally apply a selection threshold on the Critical Ratio (CR), $\rho_{\rm CR}$, computed from the FH number count map. For each source, $\rho_{\rm CR,thr}$ is set to the mean plus one standard deviation of the CR distribution, after excluding candidates overlapping known lines~\citep{known_lines_O4a} or with inconsistent detector significance. The thresholds, which are separately computed for each detector/target, range from $4.74$ to $4.87$. Detailed information on the postprocessing and follow-up stage can be found in Appendix~\ref{appendix:BSDpostprocessingFU}.

Coincidences between candidates from LIGO Hanford $(f^{\rm H}_{0\rm}, \dot{f}^{\rm H}_{0\rm})$ and Livingston $(f^{\rm L}_{0\rm}, \dot{f}^{\rm L}_{0\rm})$ are identified using the metric: 
\begin{equation} 
d=\sqrt{\left(\frac{f^{\rm H}_{0\rm}-f^{\rm L}_{0\rm}}{\delta f_{\rm FH}}\right)^2+\left(\frac{\dot{f}^{\rm H}_{0\rm}-\dot{f}^{\rm L}_{0\rm}}{\delta \dot{f}_{\rm FH}}\right)^2}, 
\label{Eq:distance_BSD} 
\end{equation}
with $d_{\rm thr}=4$ as the coincidence threshold (see \citealt{FrequencyHoughmethod}). All surviving candidates undergo a postprocessing and follow-up step.

\subsection{Postprocessing and follow-up}
\label{appendix:BSDpostprocessingFU}
After applying the known lines veto, the consistency veto and a CR threshold selection, $\sim 50$ candidates remain for each target (exact numbers are reported in Table~\ref{tab:BSDveto_summary}).

\begin{table*}[htb!]
\centering
\begin{tabular}{lcccccccc}
\toprule
\textbf{First level selection} & \textbf{G39.2} & \textbf{G65.7} & \textbf{G93.3} & \textbf{G189.1} & \textbf{G266.2} & \textbf{G291.0} & \textbf{G330.2} & \textbf{G353.6} \\
\midrule
H  & 389064 & 468268 & 468093 & 467946 & 468036 & 468170 & 468171 & 467863 \\
L  & 388524 & 467260 & 467250 & 467420 & 467375 & 467408 & 467425 &467597 \\
Coincidences & 2276 & 2728 & 2716 & 2617 & 2688 & 2634 & 2613 & 2700 \\
\bottomrule
\textbf{Veto stage}\\
\midrule
Known lines     & 1919 & 2315 & 2310 & 2204 & 2268 & 2260 & 2241 & 2291 \\
CR Consistency  & 1916 & 2311 & 2310 & 2203 & 2264 & 2255 & 2236 & 2288 \\
CR Threshold    &   39 &   55 &   61 &   44 &   47 &   42 &   44 &   36 \\
5-vector veto   & 4 & 5 & 15 & 10 & 10 & 10 & 9 & 9   \\  
Second consistency (5-vec) & 3 & 4 & 7 & 9 & 6 & 9 & 7& 8\\
Cumulative-significance veto & 1 & 1 & - & - & 3 & 1 & 1 & 1\\
\bottomrule
\end{tabular}
\caption{Number of candidates surviving the different selection criteria of the BSD post-processing stage. The top part of the table reports the first level of candidates selected from each detector (H or L) and those found in coincidence between the two. 
This is followed by the number of candidates surviving the different veto steps.}
\label{tab:BSDveto_summary}
\end{table*}

For each of these candidates, we apply a veto based on the 5-vector statistics \citep{Astone_2010,PhysRevD.89.062008}, which provides an alternative estimate of the signal-to-noise ratio and critical ratio after applying Doppler and spin-down corrections. These corrections are implemented through a heterodyne phase demodulation based on the candidate's estimated frequency $f_0$ and spin-down $\dot{f}_0$. The corrected time series is obtained by multiplying the original data by an exponential factor $\exp[-i\Phi(t)]$, where $\Phi(t)$ describes the assumed phase evolution of the signal.
A true astrophysical signal is expected to show an increase in significance after correction, scaling roughly with the fourth root of the coherence time. To test this, we compute the 5-vector statistics using two coherence times: $T_{\text{sid}} = 86164.0905$\,s (one sidereal day), and $T_{4\text{sid}} = 4T_{\text{sid}}$. For each candidate, we compute their significance $\rho^\mathrm{C}_i$ and $\rho^\mathrm{NC}_i$, where $i = \mathrm{CR}, \mathrm{snr}$ denotes either the critical ratio or the signal-to-noise ratio, and superscripts ``C'' and ``NC'' refer to corrected and uncorrected data, respectively. We retain only those candidates for which the corrected significance improves, i.e., $\rho^\mathrm{C}_i- 0.95\rho^\mathrm{NC}_i > 0$, ensuring a minimum 5\% tolerance. Simulation studies confirm that this veto has a false dismissal probability below 10\% as long as a tolerance of 5\% is assumed. At this stage, $\sim 10$ candidates per target survived, and we apply a second consistency veto using the 5-vector statistics, which further discarded some candidates. 

Among the candidates that survived these vetos, we applied the Cumulative-significance Veto, as described in Section A.1.5 of the O3a SNR paper~\citep{Abbott2021_O3aSNR}. This procedure further reduced the candidate list, eliminating the majority of remaining outliers and leaving only a few exceptions (reported in Table~\ref{tab:BSD_survivors}). 

To veto the candidates based on the Cumulative-significance Veto, we examined the trends of the critical ratio (CR), signal-to-noise-ratio, and the 5-vector S-statistics for each candidate, computed both in the corrected and uncorrected cases. Candidates that exhibited a clear reduction in significance after correction were removed, as were those with very low 5-vector CR values, specifically those below 3, corresponding to a false alarm probability bigger than $\sim1$\%. In a few cases, candidates met the first two criteria but not the third; these were tagged as low-significance outliers and retained for completeness, though flagged accordingly. The list of candidates that survived up to the Cumulative-significance Veto step is reported in Table~\ref{tab:BSD_survivors}.

\begin{table}
	\centering
	\setlength{\tabcolsep}{3pt}
	\renewcommand\arraystretch{1.2}
	\begin{tabular}{lccc}
		\toprule
		Source  & ${f_0}$ (Hz) & $\dot{f}_{0}$ (Hz/s) & $\rho_{\rm CR}$ (FH)\\
		\midrule
        G39.2$-$0.3   & 582.6910443 & $-4.07153112 \times 10^{-10}$ & 5.39 \\
		G65.7+1.2    & 155.4805487 & $-1.00577202 \times 10^{-9}$ & 5.28 \\
		G266.2$-$1.2  & 212.3495104 & $-1.61918580 \times 10^{-9}$ & 5.34 \\ 
                     & 365.5448170 & $-3.8870222216 \times 10^{-10}$ & 4.97 \\
                     & 450.3867653 & $-3.4974123560 \times 10^{-9}$ & 5.53 \\
        G291.0$-$0.1   & 563.7719252 & $-1.899359832 \times 10^{-9}$ & 5.08 \\
        G330.2+1.0   & 108.3626199 & $-7.987104272 \times 10^{-11}$ & 4.98 \\
        G353.6$-$0.7  & 30.00352801 & $-3.263776543 \times 10^{-10}$ & 6.73 \\
		\bottomrule
	\end{tabular}
	\caption{Surviving candidates from the BSD pipeline after vetos in Appendix~\ref{appendix:BSDpostprocessingFU}. The candidates investigated were excluded after comparing the corrected and uncorrected peakmap and the spectra around the candidate in each detector. For each source with a surviving associated candidate, we report the original mean CR from the FH map, the candidate frequency $f_0$, and the spin down $\dot{f}_0$ at the time of the coincidences.}
	\label{tab:BSD_survivors}
\end{table}

Following this quantitative filtering, we performed a series of visual inspections to assess the remaining candidates. This included examination of the peakmap, both before and after correction, along with the corresponding histogram, as well as inspection of the power spectral density with and without correction. These visual checks provided an additional qualitative assessment of candidate behavior under correction and helped further refine the final selection. None of the candidates was confirmed as astrophysical.

\subsection{Upper limits}
\label{appendix:BSDUL}
\begin{figure}[tbh]
    \centering \includegraphics[width=1\linewidth]{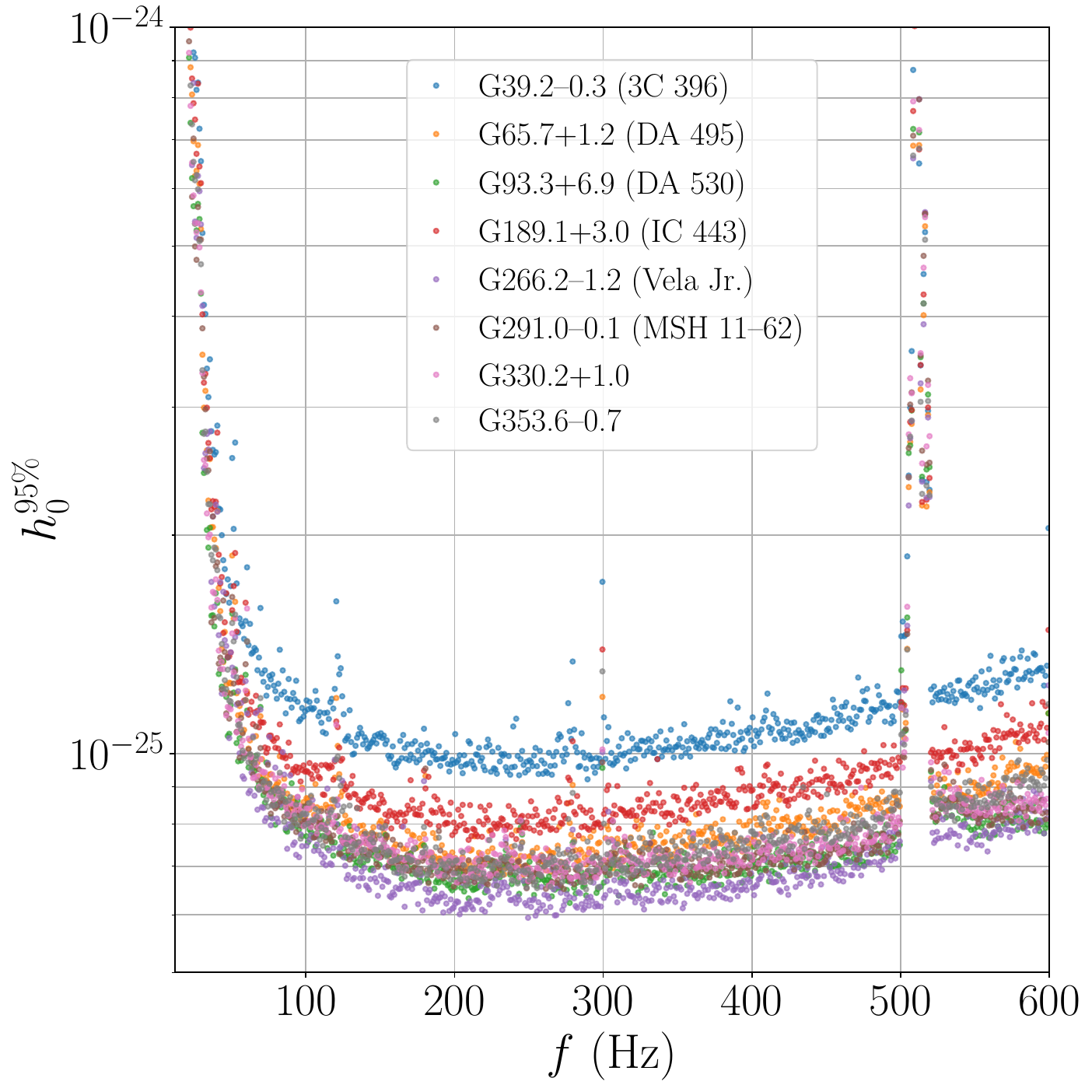}
    \caption{The estimated 95\% confidence-level upper limit obtained from the BSD search.
    }. 
    \label{fig:BSD_ULall}
\end{figure}

We set constraints on the GW strain amplitude using a well-established procedure previously applied in~\cite{DirectedBSD,O3GC}. 
The 95\% confidence level upper limit estimate is given by \citep{FrequencyHoughmethod, Palomba2025_notes_sensitivity}:
\begin{equation}
\label{eqn:h0min}
h^{95\%}_{0} \approx \frac{\mathcal{B}}{N^{1 / 4}} \sqrt{\frac{S_{n}(f)}{T_{\rm coh}}}\sqrt{ \rho_{\rm CR,max}-\sqrt{2} \operatorname{erfc}^{-1}(2 \Gamma)}
\end{equation}
where $\rho_{\rm CR,max}$ is the maximum critical ratio in each band. We indicate with $N$ the effective number of FFTs used for the search and with $\mathcal{B}$ a parameter dependent on the source sky location and the averaged signal polarization. 
The sensitivity curve, i.e. the minimum detectable strain, is obtained by substituting $\rho_{\rm CR,max}$  in Eq.~\eqref{eqn:h0min} 
with the CR threshold used for candidates selection $\rho_{\rm CR,thr}$. The resulting $h_0^{95\%}(f)$  upper limits and sensitivity curves are shown in Figure~\ref{fig:BSD_ULall}.
We corroborate the results by computing the $95\%$ confidence level upper limits on $h_0$ using artificial signals injected in randomly selected 1-Hz sub-bands: 6 sub-bands in the 10--600 Hz frequency range for all the targets. For each sub-band, $h_0^{95\%}$ is derived using a frequentist approach by injecting 50 synthetic signals per amplitude level. The injections are performed assuming the known sky position of each target. Spin-down, inclination angle ($\cos \iota$), and polarization angle ($\psi$) are drawn from uniform distributions.
We repeat the injection-and-recovery process using different values of $h_0$. 
The detection efficiency for each amplitude is then computed as the fraction of successfully recovered signals. 
A sigmoidal fit to the detection efficiency curve yields the $h_0^{95\%}$ value corresponding to 95\% detection efficiency.

To obtain a more robust estimate, the factor $\mathcal{B}$ has been calculated as the average of the empirically derived prefactors from the injections. This approach ensures that the results remain robust to potential deviations. The final upper limit curve is determined by selecting the more conservative (larger) value between the curves of each detector.

\section{Details on the PyStoch search}
\label{appendix:pystochdetails}

\subsection{Search implementation}
\label{appendix:pystochsearchtargets}
In this search, PyStoch estimates the optimal signal-to-noise ratio as a function of frequency for a specific sky location using the folded datasets as input. The targets and parameter space investigated by PyStoch are listed in Table~\ref{tab:PyStochpara}.

\begin{table*}[tbh!]
    \centering
    \begin{tabular}{lcccc}
    \toprule
       Source  & $t_{\rm age}$ (years) & $|\Delta f_0|$ (Hz) & $|\dot{f}_0|$ (Hz/s) & $\ddot{f}_0$ (Hz/s$^2$) \\
       \midrule
       G111.7$-$2.1/Cas~A  & 330 & [0.007, 3.677] & $[3.48\times 10^{-10}, 1.80\times 10^{-7}]$ & $[4.23\times 10^{-20}, 3.75\times 10^{-17}]$ \\
       G347.3$-$0.5/RX~J1713.7$-$3946  & 1600 & [0.001, 0.698] & $[6.59\times 10^{-11}, 3.41\times 10^{-8}]$ & $[1.52\times 10^{-21}, 1.35\times 10^{-18}]$ \\
       G266.2$-$1.2/Vela Jr.  & 700 & [0.003, 1.590] & $[1.51\times 10^{-10}, 7.82\times 10^{-8}]$ & $[7.98\times 10^{-21}, 7.08\times 10^{-18}]$ \\
       SNR~1987A  & 38 & [0.058, 7.460] & $[2.85\times 10^{-9}, 3.68\times 10^{-7}]$ & $[2.83\times 10^{-18}, 3.93\times 10^{-16}]$ \\
       \bottomrule
    \end{tabular}
\caption{PyStoch search setup for Cas~A, G347.3$-$0.5, Vela Jr., and SNR~1987A, assuming a given source age. The corresponding ranges of $|\Delta f_0|$, $\Dot{f}_0$, $\ddot{f}_0$ are also reported.}
\label{tab:PyStochpara}
\end{table*}

The frequencies corresponding to instrumental lines are excluded from this computation and are reported in a notch-list \citep{IsoO4a}.
The signal-to-noise ratio for each frequency bin is defined as $\frac{Y}{\sigma_{Y}}$, i.e., the ratio between the cross-correlation statistics $Y$ and the corresponding standard deviation $\sigma_{Y}$ \citep{PyStoch}.

Since GW radiometer searches are unmodeled and typically do not incorporate spin evolution, targeted narrow-band stochastic GW background analyses often employ a running average to combine the signal-to-noise ratio across adjacent frequency bins. This approach helps account for potential frequency drifts of the source over the observation period \citep{PhysRevLett.118.121102,2019PhRvD.100f2001A,2021PhRvD.104b2005A}, reducing the likelihood that signal power spreads across multiple bins and is therefore lost. To account for the frequency drift resulting from NS spin-down, we estimate the expected signal broadening, $\Delta f_0$, for each target as:
\begin{equation}
\label{Deltaf2}
    \Delta f_0  = \underbrace{-\frac{f_0}{\tage(n-1)}}_{\dot{f}_0}  T_{\text{obs}} + \frac{1}{2} \underbrace{\dfrac{n}{f_0} \biggl ( \frac{f_0}{\tage(n-1)} \biggr )^2}_{\Ddot{f}_0}  T_{\text{obs}}^2,
\end{equation}
where $T_{\text{obs}}$ denotes the duration of the O4a observing run. This spread  corresponds to the optimal frequency resolution expected for a given signal.

For the first time, we implement PyStoch with a customized binning strategy tailored to signals from young NSs in SNRs. Starting from the default 1/32 Hz resolution, adjacent frequency bins are adaptively merged to match the expected signal broadening $\Delta f_0$ for each target (see Eq.~\eqref{Deltaf2}), aiming to keep any potential CW signal fully contained within at least one frequency bin (despite possible frequency drift). To further improve sensitivity and reduce the risk of signal power being diluted across multiple bins, we apply a sliding-window combination of adjacent bins. This approach, repeated for multiple window sizes, enhances coherence detection across a range of frequency scales while preserving the original frequency grid for interpretation. Simulations presented in \cite{Nostro} indicate that combining bins beyond 1.5 Hz introduces noise clustering effects and reduces performance, establishing an empirical maximum width for effective binning.

Candidates are selected based on the simultaneous fulfillment of two criteria: (i) a signal-to-noise ratio exceeding 4.5, which corresponds to a p-value of 10 \%, and (ii) a bin width consistent with the expected signal broadening $\Delta f_0$, computed from Eq.~\eqref{Deltaf2} based on the braking indices compatible with the candidate frequency\footnote{Assume we observe an outlier at a frequency $f_0$ with a frequency distribution $\Delta f_0$ (i.e., by combining a certain number of frequency bins). If there exists an index \( n \in [2, 7] \) such that, for a given candidate frequency \( f_0 \), the corresponding value of \( \Delta f_0 \) (see Eq.~\ref{Deltaf2}) matches the expected prediction, then the pair \( (f_0, n) \) is considered compatible with the theoretical frequency distribution.}.  

The statistical significance of the outliers is quantified via $p$-values, representing the probability that a measured signal-to-noise ratio arises from random noise fluctuations. These values are estimated through Monte Carlo simulations using Gaussian realizations consistent with the noise characteristics of the data.

For each frequency bin, simulated distributions of the detection statistic are generated and processed through the same bin combination strategy described previously. The maximum signal-to-noise ratio from each realization is recorded, and the resulting ensemble is used to construct an empirical mapping between signal-to-noise ratio values and their corresponding $p$-values via linear interpolation.
Frequency bins with p-values \( \leq 10\% \) are considered statistically significant, as they are unlikely to arise from random noise fluctuations.

As a result, only a limited number of promising candidates were identified for Vela Jr. and Cas~A, while no significant candidate was observed for SNR~1987A and G347.3$-$0.5. 
These candidates were followed up using two independent methods: the Frequency-Hough transform and the 5-vector resampling technique (see Sec.~\ref{appendix:PystochFU}).

\subsection{Postprocessing and follow-up}
\label{appendix:PystochFU}
The list of candidates passed to the follow-up stage is reported in Table~\ref{tab:Pystoch_survivors}. 

\begin{table*}[!tbh]
    \centering
    \begin{tabular}{lccccccc}
    \toprule
    Source & p-value & $f_0$ (Hz) & signal-to-noise ratio & $\Delta f_0$ (Hz) & $n$ & $\dot{f_0}$ (Hz/s) & $\ddot{f_0}$ (Hz/s$^2$) \\
    \midrule
    G266.2$-$1.2/Vela Jr. & 0.09 & 1168.29688 & 4.63 & 0.34375 & 4.2 & $-1.68\times10^{-8}$ & $1.01\times10^{-18}$ \\
    G111.7$-$2.1/Cas~A     & 0.08 &  788.51563 & 4.65 & 0.59375 & 3.9 & $-2.91\times10^{-8}$ & $4.20\times10^{-18}$ \\
    G111.7$-$2.1/Cas~A     & 0.07 &  788.51563 & 4.59 & 0.71875 & 3.4 & $-3.52\times10^{-8}$ & $5.35\times10^{-18}$ \\
    \bottomrule
    \end{tabular}
    \caption{Candidates identified for each source by PyStoch. The parameters are listed from left to right as follows: $p$-value, reference frequency at the start of O4a ($f_0$), signal-to-noise ratio, candidate frequency distribution ($\Delta f_0$), braking index ($n$), and spin-down parameters evaluated at the reference time (set at the O4a start, GPS time 1368980142 s). }
    \label{tab:Pystoch_survivors}
\end{table*}

We stress that the selected candidates were subjected to follow-up analyses using two dedicated CW search pipelines: the Frequency-Hough method \citep{FrequencyHoughmethod} and the 5-vector resampling method \citep{2025CQGra..42n5008A, Sing}. No surviving candidates were confirmed by the 5-vector resampling search.
For Cas~A, the Frequency-Hough pipeline identified coincident candidates, which are defined as those lying within a distance of less than 3 in the search parameter space, when short coherence times (64 and 128 s) were used. However, these coincidences did not persist at longer coherence times (512 s), suggesting they were likely due to noise or transient features.\footnote{The Frequency-Hough follow-up analysis considered only the first-order spin-down parameter, with the second-order term being neglected. While this choice results in a reduction of the detection sensitivity (i.e., CR) for the optimal FFT length, it was a conscious choice made to simplify the initial analysis. A fully optimized version of the code, capable of incorporating and accurately handling the second-order spin-down term, will be available in the near future.} Finally, neither of the two follow-up methods identified statistically significant candidates for Vela Jr.

\subsection{Upper limits}
\label{appendix:pystochUL}
In the absence of a significant detection, upper limits on the CW strain amplitude are placed. The computation of upper limits is performed within a Bayesian framework, incorporating prior distributions over relevant source parameters such as the inclination angle, polarization angle, and calibration uncertainty. The final result is the marginalized posterior distribution for $h_0$, from which the upper limit at a given confidence level (e.g., 95\%) is extracted.

\begin{figure*}[tbh]
    \centering
\includegraphics[width=0.45\linewidth]{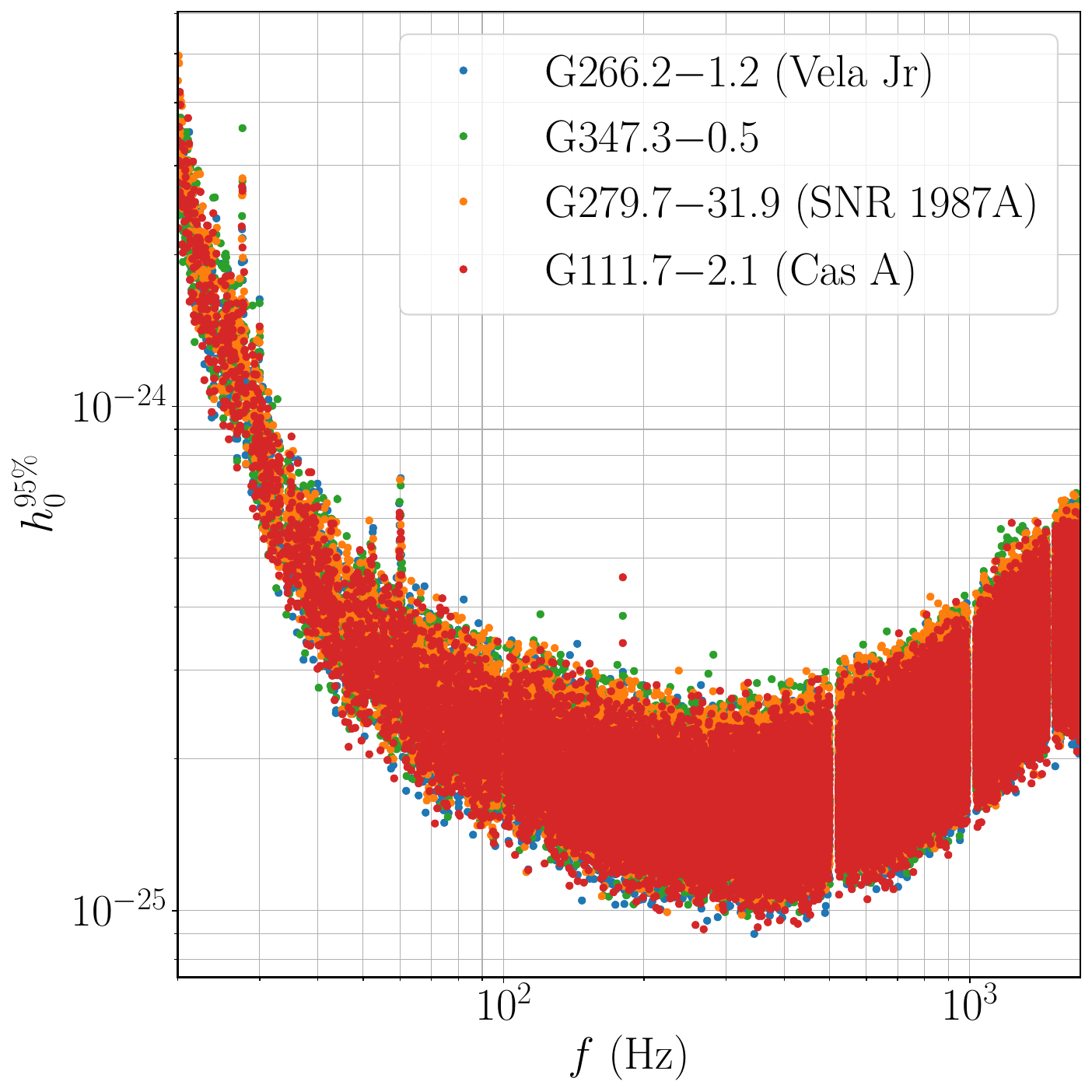}
\includegraphics[width=0.45\linewidth]{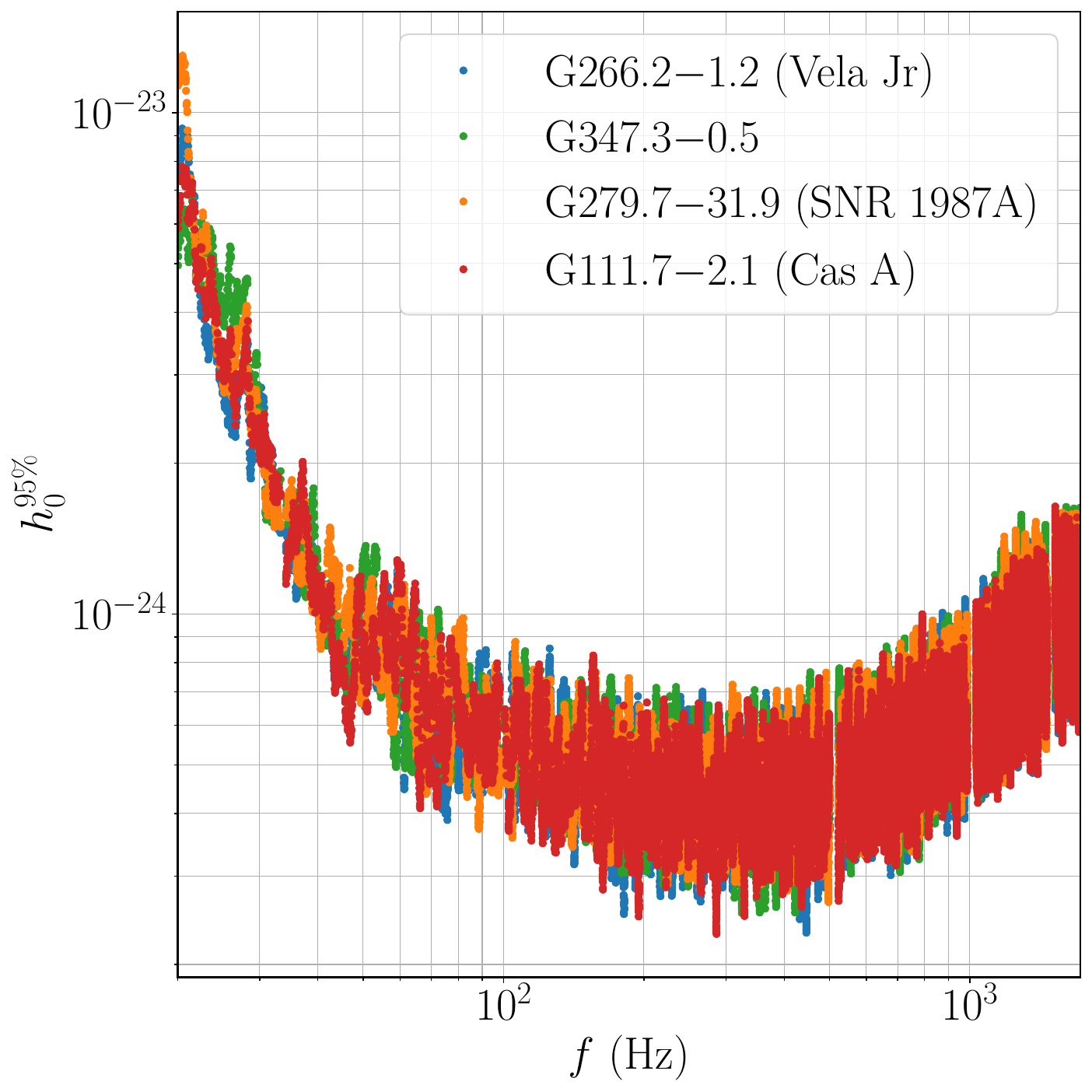}
    \caption{Best (1/32 Hz, left) and worst (47/32 Hz, right) 95\% confidence-level upper limit between 20 and 1726 Hz, computed with PyStoch for each target: SNR~1987A (orange), Vela Jr. (blue), G347.3$-$0.5 (green), Cas~A (red).}
    \label{fig:Pystoch_upperlimits}
\end{figure*}

Due to the computational cost associated with evaluating the full marginalization for each frequency bin, an interpolation-based approach is adopted. Upper limits are precomputed for a range of representative signal-to-noise ratio values and both circular and generic polarizations. The ratio between these cases, which depends only on the signal-to-noise ratio, allows for a rapid estimation of the marginalized upper limits across the entire frequency band. This procedure is validated against full posterior computations for a subset of bins.
Results for each target are reported in Fig.~\ref{fig:Pystoch_upperlimits}, while the best results are listed in Table~\ref{tab:Pystoch_uls}. 

Further technical and mathematical details of the $p$-value estimation and upper limits computation can be found in the dedicated stochastic analysis paper~\citep{StochPap}.

\begin{table}[tbh!]
    \centering
    \begin{tabular}{lcc}
\toprule
        Source & Best $h_0^{95\%}$ ($\times 10^{-26})$ & $f$ (Hz) \\
\midrule
        G111.7$-$2.1/Cas~A & $9.21 $ & 268.69 \\
        G347.3$-$0.5 & $9.78 $ & 332.53 \\
        G266.2$-$1.2/Vela Jr. & $9.02$ & 343.59 \\
        SNR~1987A & $9.83$ & 323.16 \\
        \bottomrule
    \end{tabular}
    \caption{Best 95\% confidence-level upper limit set by PyStoch with corresponding frequency for each SNR.}
    \label{tab:Pystoch_uls}
\end{table}

\section{Details on Single-harmonic Viterbi search}
\label{appendix:SHV_details}
\subsection{Search implementation}
\label{appendix:SHV_searchtargets}

In this appendix, we summarize the SHV pipeline used in Sec. \ref{sec:shv}.
We refer the reader to \cite{bu_ligot2500368v1_2025} which includes the technical and full details of this method and major update to the SHV pipeline since the O3a search \citep{Abbott2021_O3aSNR}.

We follow the Viterbi Hidden Markov Model  (HMM) formalism used in previous SHV searches (e.g., \citealt{suvorova_hidden_2016, sun2018hidden}). We consider a set of $N_T$ observations at discrete times $t_0, t_1, \ldots, t_{N_T - 1}$ and $N_Q$ possible hidden states $\left\{ q_1, \ldots, q_{N_Q } \right\}$, where each hidden state corresponds to a discrete GW frequency bin. 
The corresponding sequence of discrete states $\left\{q(t_0), q(t_1), \ldots, q(t_{N_T - 1}) \right\}$, through which the system progresses, forms a Markov chain. 
The transition probability from time $t_k$ to $t_{k+1}$ is defined as $A_{q_i q_j} = P\left[q(t_{k+1}) = q_j \mid q(t_k) = q_i\right]$.
We adopt a simple random-walk transition model in frequency, allowing the signal to remain in the same bin or wanders up or down by one bin, during each time step with equal probability.
We assume a uniform prior over the initial hidden state.
The emission probability is computed from the coherent $\mathcal{F}$-statistic in each segment of duration $T_{coh}$. Given the observation sequence, the most likely hidden state sequence
can be inferred by maximizing
the most probable hidden-state path.

\subsubsection{Parameter choice}
\label{appendix:SHVparameter_choice}

Here we outline the recipe for selecting the searchable parameter domain in the SHV pipeline. The list of targets and the parameter domain for each target are listed in Table~\ref{tab:SHVvit_targets}.
For each target with distance $D_{\textrm{min}} \leq D \leq D_{\textrm{max}}$, age $t_{\textrm{age, min}} \leq \tage \leq t_{\textrm{age, max}}$, and braking index $n\smin \leq n \leq n\smax$, we define the search parameters $\Tcoh$ and $f$ via four criteria:
\begin{enumerate}
    \item The frequency bin width $\delta f$ is sufficiently small, such that $f$ remains constant during each coherent segment, with 
    \begin{equation} \label{eq:deltaf}
        \delta f = \frac{1}{2\Tcoh} > |\dot{f}\left(f, n \right)| \Tcoh.
    \end{equation}
    \item The $\mathcal{F}$-statistic software ingests at least two SFTs \citep{LSC_sft_2022}, with $\Tcoh \geq 2T_{\rm SFT}$.
    \item The signal amplitude $\hsig$ exceeds the minimum detectable strain at 95\% confidence for a semicoherent search $\hest$, with $\hsig > \hest$.
    \item The maximum strain implied by energy conservation, $\hmax$, exceeds the signal amplitude, with $\hmax > \hsig$. 
\end{enumerate}

\begin{table*}[ht!]
\centering
\begin{tabular}{lcccc}
\toprule
Source & $f$ (Hz) & $\Tcoh$ (hours) & $\Tcoh$ (hours) & $\dot{f}$ ($\mathrm{Hz\,s^{-1}}$) \\ 
& & & (@100 Hz) & (@100 Hz) \\
\midrule
G18.9$-$1.1            & [29, 2000]   & [4.7, 39.2]   & 21.1   & [$-8.67\times 10^{-11}$, $8.67\times 10^{-11}$] \\
G39.2$-$0.3/3C396      & [55, 736]    & [7.1, 26.3]   & 19.5   & [$-1.01\times 10^{-10}$, $1.01\times 10^{-10}$] \\
G65.7+1.2/DA495       & [21, 2000]   & [23.7, 228.7] & 106.4  & [$-3.41\times 10^{-12}$, $3.41\times 10^{-12}$] \\
G93.3+6.9/DA530       & [39, 1379]   & [6.0, 36.3]   & 22.6   & [$-7.55\times 10^{-11}$, $7.55\times 10^{-11}$] \\
G189.1+3.0/IC443      & [21, 2000]   & [10.4, 102.2] & 46.8   & [$-1.76\times 10^{-11}$, $1.76\times 10^{-11}$] \\
G266.2$-$1.2/Vela Jr.   & [20, 2000]   & [4.3, 43.2]   & 19.3   & [$-1.04\times 10^{-10}$, $1.04\times 10^{-10}$] \\
G353.6$-$0.7           & [30, 2000]   & [9.9, 80.7]   & 44.4   & [$-1.96\times 10^{-11}$, $1.96\times 10^{-11}$] \\
G1.9+0.3              & [32, 1607]   & [1.0, 7.1]    & 4.0    & [$-2.41\times 10^{-9}$,  $2.41\times 10^{-9}$] \\
G15.9+0.2             & [34, 1915]   & [3.3, 25.4]   & 14.8   & [$-1.76\times 10^{-10}$, $1.76\times 10^{-10}$] \\
G111.7$-$2.1/Cas~A     & [28, 2000]   & [1.1, 9.5]    & 5.1    & [$-1.48\times 10^{-9}$,  $1.48\times 10^{-9}$] \\
G291.0$-$0.1/MSH11-62  & [30, 2000]   & [6.0, 48.9]   & 27.0   & [$-5.29\times 10^{-11}$, $5.29\times 10^{-11}$] \\
G330.2+1.0            & [31, 2000]   & [7.4, 59.2]   & 33.1   & [$-3.52\times 10^{-11}$, $3.52\times 10^{-11}$] \\
G347.3$-$0.5           & [21, 2000]   & [2.4, 24.2]   & 11.1   & [$-3.13\times 10^{-10}$, $3.13\times 10^{-10}$] \\
G350.1$-$0.3           & [32, 1999]   & [1.5, 12.7]   & 7.2    & [$-7.44\times 10^{-10}$, $7.44\times 10^{-10}$] \\
\bottomrule
\end{tabular}
\caption{Sources searched in the SHV analysis (Section \ref{sec:shv}) with their search bands and coherence times $\Tcoh$, and the corresponding frequency derivative ranges $\dot{f}$ computed as $1/(2T_{\mathrm{coh}}^2)$ using the @100 Hz coherence time (converted to seconds). 
The age and distance of each target are listed in Table \ref{tab:targets}. 
SHV does not specifically search the first-order spin-downs. The estimated spin-down reported is inferred from $\Tcoh$.}
\label{tab:SHVvit_targets}
\end{table*}

\begin{figure}[htbp]
    \centering
    \includegraphics[width=1\linewidth]{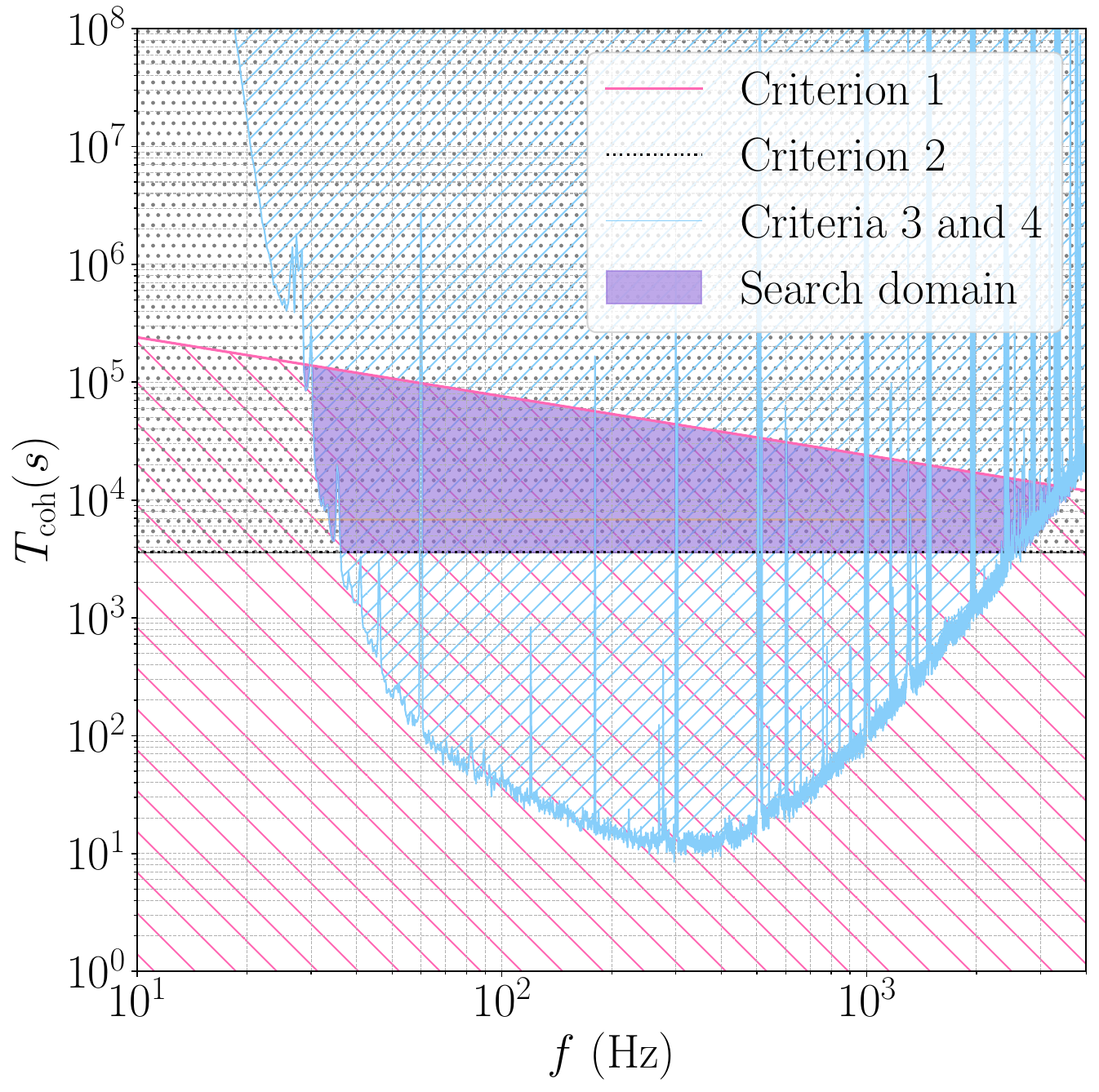}
    \caption{
         $\Tcoh$ and $f$ values that satisfy the four criteria defined in Appendix \ref{appendix:SHVparameter_choice}. 
        The shaded pink region satisfies criterion one, \ie \Eq \eqref{eq:td_c1}.
        The dotted grey region satisfies criterion two. 
        The shaded blue region satisfies criteria three and four simultaneously.
        The purple domain indicates where all four criteria are satisfied, designated as the search region. 
        For each $f$ in the purple region, we choose $\Tcoh$ according to the pink curve to maximize the search sensitivity. 
    }
    \label{fig:SHVparam_space}
\end{figure}

Using Eq.\ \eqref{eq:powlaw} to express $\dot{f}$ and adopting the standard late-time approximation 
\begin{align}\label{eq:fdot}
\dot{f}(t_{\rm age}) &\approx - \frac{f(t_{\rm age})}{(n-1) \tage,} 
\end{align}
the first criterion yields an upper bound on coherence time:
\begin{align}\label{eq:td_c1}
    \Tcoh &< \left[ \frac{2f}{(n-1)\tage} \right] ^{-1/2}
\end{align}
The second criterion provides a lower bound on $T_{coh}$. The third and fourth criteria are simultaneously satisfied when $\hmax > \hest$, which gives a lower bound of the form
\begin{align} \label{eq:3_and_4}
    \Tcoh > &\left(2.27 \times 10^{-24} \, \mathrm{kpc} \, \mathrm{kyr}^2\right) ^{-4} \ \ \nonumber \\ & D ^4 \ \tage ^{2} (n-1) ^{2} \Theta ^ 4\ S_n\left(f\right)^{2} \ \Tobs ^{-1}, 
\end{align}
where $\Theta = 35$ is an empirical statistical factor \citep{wette_searching_2008}.
The resulting overlap of constraints defines the searchable ($f$, $ \Tcoh$) domain. For each frequency within this domain, we choose $ \Tcoh$ to maximize sensitivity, following the prescription illustrated in Fig.\ \ref{fig:SHVparam_space} and summarized in Table~\ref{tab:SHVvit_targets}.

\subsubsection{Detection Threshold}
\label{appendix:SHVdetection_threshold}
We set detection thresholds by choosing an overall false alarm probability, $\alpha = 0.30$, for the ensemble of 14 targets. Across all targets, with each labelled by $i$, there are $N_{\text{band}, i}$ sub-bands.
Hence the equivalent false alarm probability per sub-band for target $i$ is $    \alpha_{\text{band}, i} = 1 - (1 - \alpha_{\rm target}) ^{1 /  N_{\text{band}, i}}$.

Following previous SHV searches, we determine the threshold for the Viterbi log-likelihood statistic $\mathcal{L}$ via Monte Carlo simulations in Gaussian noise with the O4a duty cycle. The likelihood threshold $\Lth$ is defined as the $1- \alpha_{\text{band}, i}$-th percentile
of the $\mathcal{L}$ distribution. The resulting thresholds are well described by a power-law dependence on coherence time,
$\Lth = C_{\rm band} \Tcoh ^{p_{\rm band}}$
which we use to evaluate thresholds across the search space. Any path with $\mathcal{L} > \Lth$ is followed up using the procedure in Appendix 
\ref{appendix:SHV_followup}. For a detailed analysis of the individual targets, see \cite{bu_ligot2500368v1_2025}.

\subsection{Postprocessing and follow-up}
\label{appendix:SHV_followup}
After conducting the SHV search, we apply a standard follow-up procedure comprising a known-line veto, an interferometer veto, a sky-position veto, and a sub-band veto (see \citealt{Abbott2021_O3aSNR}). The number of candidates remaining after each stage is summarized in 
Table~\ref{tab:SHV_veto}.
A total of six candidates survive all automated vetoes; their parameters are listed in Table~\ref{tab:SHV_followup}.

\begin{table*}[ht]
\centering
\begin{tabular}{lcccccccccccccc}
\toprule
\textbf{Veto stage} & \textbf{G111} & \textbf{G65} & \textbf{G1} & \textbf{G291} & \textbf{G266} & \textbf{G189} & \textbf{G330} & \textbf{G353} & \textbf{G347} & \textbf{G18} & \textbf{G93} & \textbf{G350} & \textbf{G15} & \textbf{G39} \\
\midrule
Above threshold    & 435 & 368 & 317 & 316 & 313 & 294 & 294 & 293 & 265 & 264 & 261 & 258 & 213 & 109 \\
Known‐lines veto    &  31 &  60 &  27 &  29 &  30 &  33 &  37 &  28 &  24 &  22 &  25 &  21 &  25 &   8 \\
Interferometer veto &   1 &  14 &   1 &   1 &   3 &   4 &   2 &   1 &   2 &   2 &   2 &   1 &   1 &   - \\
Sky position veto   &   1 &  14 &   1 &   1 &   3 &   4 &   1 &   1 &   1 &   2 &   2 &   - &   1 &   - \\
Sub‐band veto       &   1 &   3 &   1 &   - &   1 &   - &   - &   - &   - &   - &   - &   - &   - &   - \\
\bottomrule
\end{tabular}
\caption{
Survivor counts for each target of SHV with $\mathcal{L} > \Lth$ and at the end of all four post-processing stages: known‑lines veto, interferometer veto, sky position veto, sub‑band veto. We shorten the names of the targets for simplicity as G347 for G347.3$-$0.5, etc. The full names are reported in Table~\ref{tab:targets}.
    }
\label{tab:SHV_veto}%
\end{table*}

\begin{table*}[ht]
\centering
\begin{tabular}{l c c c c c c}
\toprule
Source & $f_0$ (Hz) & $\Lth$ & $\mathcal{L}$ & $\mathcal{L}_{\rm H1}$ & $\mathcal{L}_{\rm L1}$ & $\mathcal{L}_{\rm off-target}$ \\
\midrule
G111.7$-$2.1/Cas~A    & 1169.00 & 27861.43 & 35500.86 & 36564.78 & 42308.51 & 35078.42 \\
G266.2$-$1.2/Vela Jr. &  213.31 &  3177.03 &  3777.69 &  3339.62 &  5149.54 &  8557.62 \\
G65.7+1.2/DA 495    &  108.76 &   418.83 &   430.26 &   442.83 &   419.67 &   432.77 \\
G65.7+1.2/DA 495    &  137.83 &   472.39 &   483.77 &   474.44 &   477.31 &   478.21 \\
G65.7+1.2/DA 495    &  139.95 &   474.62 &   491.81 &   568.67 &   484.56 &   545.42 \\
G1.9+0.3           & 1161.08 & 34800.84 & 45007.50 & 54777.89 & 35121.50 & 45144.90 \\
\bottomrule
\end{tabular}
\caption{
    Candidates surviving all vetos from the SHV search. 
    Each row lists a candidate’s target name, start frequency $f_0$ of the Viterbi path, threshold log‑likelihood $\mathcal{L}_{\rm th}$, combined detection statistic $\mathcal{L}$, single-interferometer log‑likelihoods $\mathcal{L}_{\rm H1}$ and $\mathcal{L}_{\rm L1}$ from the interferometer veto, and off‑target log‑likelihood $\mathcal{L}_{\rm off-target}$ from the sky‑position veto.}
\label{tab:SHV_followup}
\end{table*}

We examine these survivors using single-interferometer statistics and off-target sky maps. None display the expected on-target morphology or cross-detector consistency of a true signal. We therefore conclude that all six outliers are consistent with instrumental artifacts and that no SHV candidate survives follow-up.

\subsection{Upper limits}
\label{appendix:SHV_UL}
We determine the upper limits using all targets. 
For each target, we choose two sub-bands within the original search that have no known lines and return fewer than two unique paths with $\mathcal{L} > \Lth$, with $f \simeq f_{\min}$ and $f \simeq f_{\max}$, where $f_{\min}$ and $f_{\max}$ denote the minimum and maximum frequency of the target's search band. 
The chosen frequencies are shown in~\cite{bu_ligot2500368v1_2025}, along with $\Tcoh$ and $\Lth$ from the original search. 

\begin{figure}[htbp]
  \centering
  \includegraphics[width = 1\linewidth]{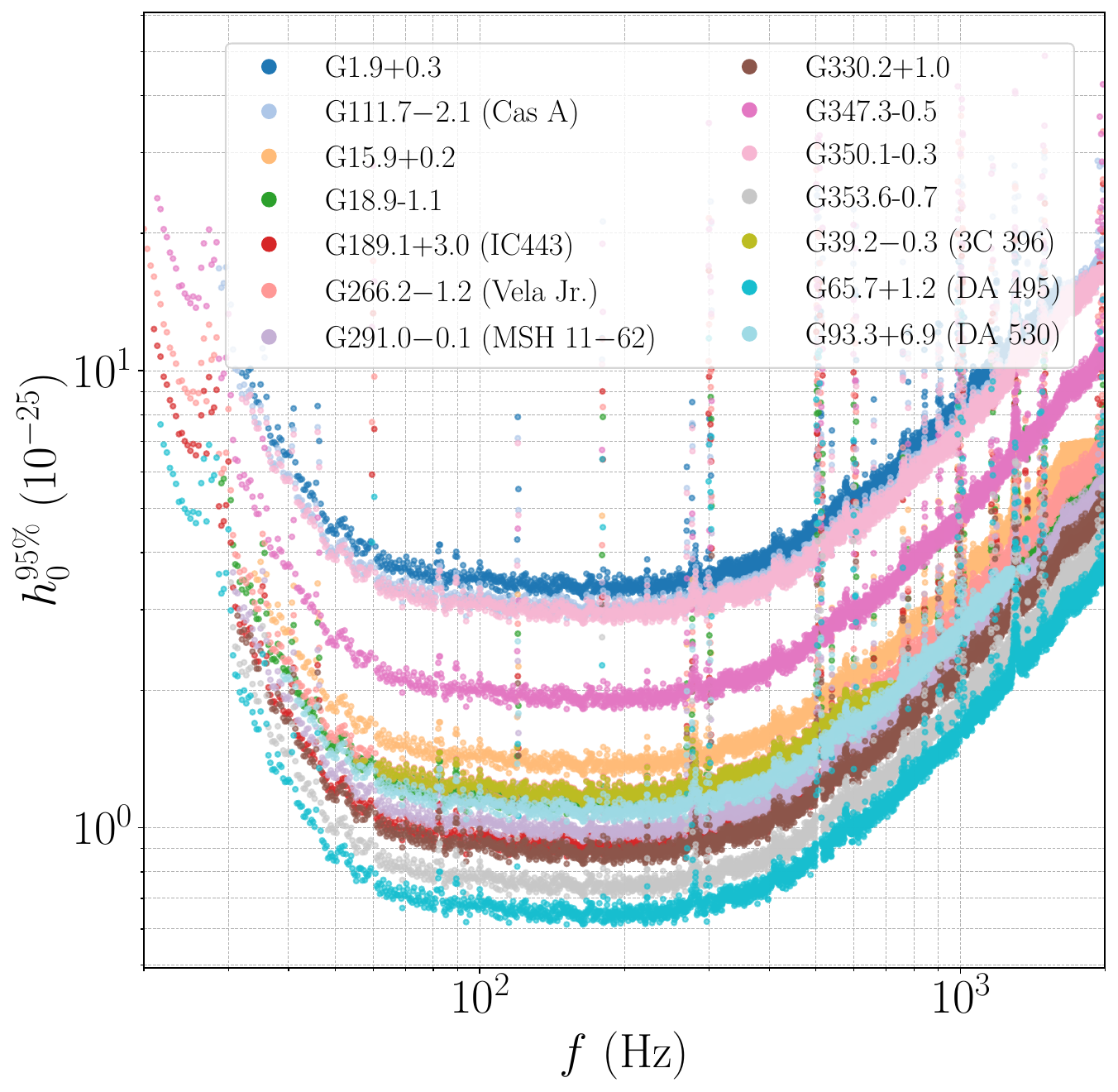}
  \caption{Upper limit curves for all targets searched by the SHV pipeline. The scatter points show the sensitivity estimate  $h_0^{95\%}(\Tcoh) = a(\Tcoh) \hest$ across the searched frequency band.}
  \label{fig:h_95_SHV}
\end{figure}

By recording the fraction of injected signals recovered (\ie those with $\mathcal{L}>\Lth$) for each $h_0$, we estimate the detection efficiency $\epsilon(h_0)$. 
We then use logistic regression to fit a sigmoidal model to $\epsilon(h_0)$, and solve
$\epsilon\bigl(h_0^{95\%}\bigr) = 0.95$ to obtain the 95\% confidence upper limit $h_0^{95\%}$ in the given sub-band.
We calculate $a = h_0^{95\%} / \hest$ for each sub-band. 
We assume that the sensitivity only depends on $\Tcoh$ and neglect other factors such as sky position.
To obtain the upper limit across the full search band, we define $h_0^{95\%}(\Tcoh) = a(\Tcoh)\,\hest$.
To verify that the full-band scaled upper limit $h_{0,\mathrm{UL}}^{95\%}$ is reliable, we compare it with the upper limit obtained from injections $h_{0,\mathrm{inj}}^{95\%}$.
For each target and each injection, we compute the ratio $R \equiv {h_{0,\mathrm{inj}}^{95\%}}/{h_{0,\mathrm{UL}}^{95\%}}$.
For each target, we then define a rescaling constant per target $k = \max(R) \quad \text{over injections for that target}$.
If $k > 1.1$, the entire upper-limit curve for that target is rescaled as $h_{0,\mathrm{UL}}^{95\%} \rightarrow k\, h_{0,\mathrm{UL}}^{95\%}$.
See \cite{bu_ligot2500368v1_2025} for the values of $k$ for each target.
We present the upper-limit curve for all targets in \Fig\ \ref{fig:h_95_SHV}.

The SHV pipeline models the frequency evolution as a random walk, which is flexible enough to accommodate secular spin down and stochastic spin wandering.
Consequently, the resulting upper limits are not directly comparable to the pipelines in this paper which model the frequency evolution as a Taylor series.

\section{Details on the Dual-harmonic Viterbi search}
\label{appendix:dhv_details}
\subsection{Search implementation}
\label{appendix:dhv_detailssearchtargets}
The HMM structure used in the dual-harmonic search largely follows the standard formulation described in Appendix~\ref{appendix:SHV_details}, with specific modifications outlined in this Appendix. 
The list of targets and parameter space covered is listed in Table~\ref{tab:DHVpara}.

\begin{table*}[!tbh]
	\centering
	\setlength{\tabcolsep}{5pt}
	\renewcommand\arraystretch{1.2}
	\begin{tabular}{lcccc}
		\toprule
		Source & Minimum $\tage$ (kyr) & $T_{\rm coh}$ (hr) & $f_\star$ (Hz)& $\dot{f}_\star$ $(\rm{Hz/s})$ \\
		\midrule
		G353.6$-$0.7 & 20 & 12 &  [50, 338]& [$-1.34\times 10^{-10}$, 0]  \\
		G189.1+3.0 &&&&  \\
		G65.7+1.2 &&&&  \\
        \hline
		  G266.2$-$1.2& 5 & 9 & [50, 150] & [$-2.38 \times 10^{-10}$, 0] \\
		G93.3+6.9  &&&& \\
		  G18.9$-$1.1 &&&&  \\
        G39.2$-$0.3  &&&&  \\
		\bottomrule
	\end{tabular}
	\caption{Targets searched in the dual-harmonic Viterbi analysis (Section~\ref{appendix:dhv}) and the parameter space covered.}
	\label{tab:DHVpara}
\end{table*}

In this analysis, two frequency components are tracked simultaneously. The signal power in the $i$th frequency bin, $f_i$, is computed by the sum of two $\mathcal{F}$-statistic components computed in dual-harmonic frequency bins, $\mathcal{F}_1(f_i) + \mathcal{F}_2(2f_i)$.
The GW signal for each frequency component is assumed to be monochromatic over $\Tcoh$.

We choose a coherence time interval $\Tcoh$ such that the following condition is satisfied:
\begin{equation}
\label{eqn:int_T}
\left|\int_t^{t+\Tcoh}dt' \dot{\fstar}(t')\right| < \delta f_{1},
\end{equation}
where $\delta f_{1} = 1/(4\Tcoh)$ is the frequency bin width in the $\mathcal{F}_1$ statistic. For the $\mathcal{F}_2$ statistic, we adopt a bin width of $\delta f_{2}=2\delta f_{1} = 1/(2\Tcoh)$. The choice of bin widths ensures that the signal drifts by no more than one bin in both $\mathcal{F}_1$ and $\mathcal{F}_2$ across adjacent coherent segments.
Thus, the $\dot{f}_\star$ range covered in the search (listed in Table~\ref{tab:DHVpara}) depends on $\Tcoh$, i.e., $|\dot{f}_\star| \in [0, 1/(4\Tcoh^2)]$.

The emission probability for each time step of length $\Tcoh$ is computed as \citep{Jaranowski1998, Sun2019}:
\begin{eqnarray}
\label{eqn:emi_prob_matrix}
\nonumber
\ln L_{o(t_k) q(t_k)} &=& \ln P[o(t_k) \mid f_i \leq \fstar(t_k) < f_i + \delta f_{1}] \\
&=& \mathcal{F}_1(f_i) + \mathcal{F}_2(2f_i),
\end{eqnarray}
where $f_i$ denotes the frequency in the $i$-th bin.

In this work, we assume that the frequency evolution is dominated by the secular spin down of the NS. Under this assumption, the transition probability matrix $A_{q_j q_i}$ reduces to the form \citep{Sun2019}:
\begin{equation}
\label{eqn:trans_matrix_sd}
A_{q_{i-1} q_i} = A_{q_i q_i} = \frac{1}{2},
\end{equation}
with all other entries set to zero. A uniform prior distribution is applied to the initial state, $\Pi[q(t_0)] = N_Q^{-1}$.

The Viterbi score $S$ is used as the detection statistic in this pipeline, representing the significance of the optimal Viterbi path in each sub-band relative to all other paths at the final step of the tracking~\citep{Sun2019}.

Two different coherence times of $\Tcoh = 12$~hr and 9~hr are selected for three older sources and four younger sources, respectively (see Table~\ref{tab:DHVpara}). For parallelized computing, the full frequency band is divided into 1-Hz (1.5-Hz) sub-bands for $\Tcoh = 12$~hr ($\Tcoh = 9$~hr).\footnote{The detection statistic used in this analysis requires a much larger number of frequency bins in each sub-band than the total number of tracking steps (i.e., $2B\Tcoh \gg \Tobs/\Tcoh$ with $B$ being the bandwidth), such that the paths ending in different bins are mostly independent. Therefore, we extend the sub-band to 1.5~Hz for $\Tcoh = 9$~hr to satisfy this requirement.}

We set detection thresholds corresponding to 1\% false alarm probability at $S_{\rm th}=5.26$ for $\Tcoh = 12$~hr and $S_{\rm th}=5.15$ for $\Tcoh = 9$~hr, based on Monte-Carlo simulations in Gaussian noise and validated using real O4a data. 

After conducting the search for each target and identifying all the above-threshold candidates, we apply a standard follow-up procedure with a series of vetoes described in Sec.~\ref{appendix:dhv_final_veto}.

\subsection{Postprocessing and follow-up}
\label{appendix:dhv_final_veto}
After a standard procedure of the known-line veto, interferometer veto, Doppler-shift veto, and sky-position veto (see details in \citealt{Abbott2021_O3aSNR}), 16 candidates remain, listed in Table~\ref{tab:viterbi1f2f_survivors}. We pass these 16 survivors through the final stage of a coherence-time veto, and all of them are rejected. 
With the increased $\Tcoh$ listed in the sixth column, all except one of the new scores fall below the original score. 
To be more conservative, we further check the two candidates with a new score below the original score but above $\Sth$, marked by ``*'' and ``$\ddagger$''. 
With the increased $\Tcoh$, a lower mean spin-down rate is returned for both candidates, allowing us to further increase the coherence time to $\Tcoh = 20$~hr. We find that the score further drops for both candidates. We conclude that they do not survive further scrutiny. 
One candidate, marked by ``$\dagger$'', returns with a higher score of 14.38, but the optimal path (at $\approx 260.79$~Hz, intersecting a known instrumental line at Hanford) is distinct from the original one. By shifting the search sub-band to 259.7$-$260.7~Hz with $\Tcoh = 15$~hr, the score drops below $\Sth$, and the new optimal path does not match the original. 
Thus, none of the remaining candidates survives the final check. 

\begin{table*}[!tbh]
	\centering
	\setlength{\tabcolsep}{3pt}
	\renewcommand\arraystretch{1.2}
	\begin{tabular}{lccccccc}
		\toprule
		Source & $S$ &$2{f_\star}_{\rm start}$ (Hz) & $2{f_\star}_{\rm end}$ (Hz)& Mean $2\dot{f}_\star$ ($10^{-10}$~Hz/s) & New $\Tcoh$ (hr) & New $S$ & Paths overlap?\\
		\midrule
		G353.6$-$0.7  & 6.68   &	184.9413888893380 &	184.9388657411880 &	$-$1.23198& 15& 6.04 * & $\checkmark$\\
		& 5.58   &   346.2777314809320 & 346.2748495364930 &	$-$1.40717  & 15 & 4.47& $\times$\\
		& 5.32   &	586.6338310172640 &	586.6316666654170 &	$-$1.05679  & 15& 5.01 & $\checkmark$\\
		& 5.59 & 	641.9995486091329 &	641.9969907387680 &	$-$1.24894 & 15 & 4.52 & $\times$\\
        \hline
		G189.1+3.0 & 6.10&260.6886458319710 &	260.6859606467910 &	$-$1.31110 &  15& 14.38 $\dagger$& $\times$\\
        & 6.16	&272.0795138887320 &	272.0770949072550 &	$-$1.18112 &  15& 5.77 $\ddagger$& $\checkmark$\\
        & 5.53	&382.4306828695180 &	382.4274074065610 &	$-$1.59932 &  15& 4.49& $\times$\\
        & 5.62	&405.6810763875410 &	405.6783564801390 &	$-$1.32805 &  15& 4.49& $\checkmark$\\
        & 5.28	&431.0656134257960 &	431.0625810183950 &	$-$1.48064 &  15& 5.02& $\checkmark$\\
        & 6.37	&498.9999999980210 &	498.9982060165430 &	$-$0.87595 &  15& 4.47& $\times$\\
        & 5.41	&608.6361921283710 &	608.6331018505990 &	$-$1.50890 &  15& 4.86& $\times$\\
        \hline
		G65.7+1.2  & 5.90   & 396.2719791661280 &	396.2690393513190 &	$-$1.43543 & 15& 4.17& $\times$\\
		&  5.38   & 412.6593865727690 &	412.6568055542560 &	$-$1.26024 & 15 &4.13& $\times$\\
		&  5.56   & 441.6801273134690 &	441.6772337949560 &	$-$1.41282 & 15 &4.00& $\times$\\
		&  5.27   & 575.3298726845320 &	575.3267361104640 &	$-$1.53150 & 15  &3.97& $\times$\\
        \hline
		G18.9$-$1.1 &	5.28   & 117.3951234567360 &	117.3906481480940 &	$-$2.18517 &12 & 3.94& $\times$\\
		\bottomrule
	\end{tabular}
	\caption{Final candidates from the dual-harmonic Viterbi pipeline and the coherence-time veto results (all vetoed). The first five columns list the source name, original score, estimated start and end 2$f_\star$, and the mean $2\dot{f}_\star$. Column 6 lists the new $\Tcoh$ used in the coherence-time veto. The last two columns show the follow-up results: the new score obtained by increasing $\Tcoh$, and whether the new optimal path overlaps the original candidate path. The three new scores marked by ``*'', ``$\dagger$'', and ``$\ddagger$'' are above the original thresholds set in the search.Note that the pipeline returns estimated frequencies and spin-down rates corresponding to the $2f_\star$ component.}
	\label{tab:viterbi1f2f_survivors}
\end{table*}

\subsection{Upper limits}
\label{appendix:DHV_UL}
\begin{figure}
	\centering
    \includegraphics[width = 1\linewidth]{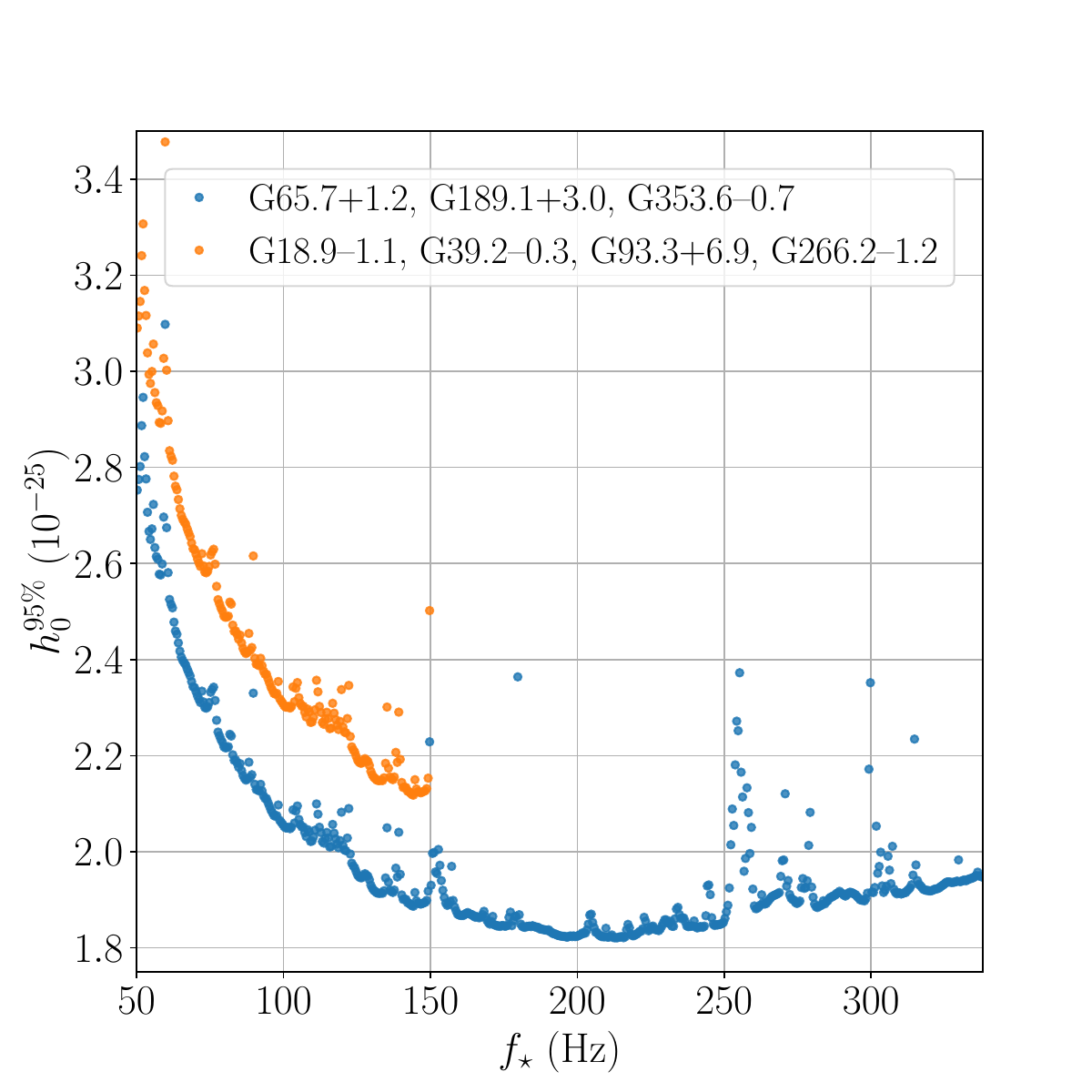}
	\caption{The estimated sensitivity $h_0^{95\%}$ obtained from the dual-harmonic Viterbi search as a function of $\fstar$ for $\Tcoh=12$~hr (blue) and $\Tcoh=9$~hr (orange), assuming a specific scenario with source properties $\theta=45$~deg and $\cos\iota = 0$ (signals at both $\fstar$ and $2\fstar$ are linearly polarized). 
	These estimates are obtained from randomized sky positions and hence apply to all sources using the same $\Tcoh$. 
    }
	\label{fig:viterbi1f2fUL}
\end{figure}

In Fig.~\ref{fig:viterbi1f2fUL}, we show the upper limits for the DHV targets. 
First, we derive the frequentist upper limit through injections in one sample sub-band, starting from 112.5~Hz for $\fstar$ and 225~Hz for $2\fstar$. A set of 200 synthetic signals is injected into the O4a data at random sky positions in that sub-band with a fixed $h_0$.  
The corresponding detection rate is calculated. 
This process is repeated with different $h_0$ values with step size $1 \times 10^{-26}$. With all the injected $h_0$ values and the corresponding detection rates, $h_0^{95\%}$ is obtained through a sigmoidal fit. 
We find $h_0^{95\%} = 2.04 \times 10^{-25}$ and $2.29 \times 10^{-25}$ for $\Tcoh=12$~hr and 9~hr, respectively, in this sample band.
The $h_0^{95\%}$ values obtained in the sample sub-band improve upon the O3a results by a factor of approximately 1.4~\citep{Abbott2021_O3aSNR}. Given that the detectable strain amplitude scales with the effective amplitude spectral density and inversely with the square root of the observation time, the sensitivity achieved in O4a is consistent with expectations based on the O3a results.

Next, we use these values obtained in the sample sub-band (starting from 112.5~Hz for $\fstar$ and 225~Hz for $2\fstar$) to analytically calculate $h_0^{95\%}$ in the full frequency band (blue dots), using the scaling~\citep{Sun2019}
\begin{equation}
h_0^{95\%}(f) \propto \left(\frac{S_n(f)S_n(2f)}{S_n(f)+S_n(2f)}\right)^{1/2},
\label{eq:ULscaling}
\end{equation}
where $S_n$ is the effective power spectral density calculated from the harmonic mean of the two detectors over all the 30-min SFTs collected from October 1--31, 2023 (GPS time 1380153618--1382745618).
Finally, to verify the analytical scaling, the injection procedure in the first step is repeated in 12 randomly selected sub-bands that span the full range. 
The empirically obtained $h_0^{95\%}$ in these sample sub-bands agree with the analytic estimates to $< 2.1 \%$.

\section{Details on the Weave search} \label{appendix:weave_pipelinesdetails}
\subsection{Search implementation}
\label{appendix:weave_searchtargets}
The Weave software provides a systematic way to cover the parameter space in a template search, ensuring an acceptable loss of signal-to-noise ratio for true signals that fall between template points~\citep{Wette2018}. It combines recent advances in template placement, using an optimal parameter-space metric~\citep{Wette2013,Wette2015} and optimal template lattices~\citep{Wette2014}.

Due to the high computational cost of template-based Weave searches, we limit our search to \casa, \vela, and \targetg for GW signal frequency $f$ ranges from 20 to 475 Hz. 
Detector noise deteriorates badly below 20 Hz because of ground motion and is also badly degraded at much higher frequencies in bands contaminated by forests of ``violin mode'' frequencies, near 500 Hz and near integer multiples thereof. These frequencies correspond to vibration resonances of the 16 suspension fibers supporting the four primary mirrors in each interferometer.
Similar but much less severe resonances in suspension fibers supporting the beam splitter mirrors contaminate narrow bands between 300 and 400 Hz.
In this search we focus on frequencies below the fundamental violin mode harmonics for two reasons: 1) young, but not newborn, isolated NSs are expected to rotate at lower frequencies; and 2) higher-frequency searches are computationally more demanding because of the increase in parameter space volume. 
We choose a ceiling of 475 Hz because the region between 475 Hz and the lowest violin mode harmonic (just below 500 Hz) is contaminated by non-linear interactions between the violin modes and other, lower-frequency lines, leading to frequency ``shoulders'' and sidebands.

Similar to previous searches~\citep{abadie2010first, aasi2015searches, Abbott_2019, Abbott2021_O3aSNR, O3aSNRweave, fullo3_casA, 2026arXiv260304792C}, we define the search range by assuming a power-law spin-down: $\dot{f} \propto -f^n$, with a braking index $n$ ranging between 2 and 7 to account for multiple potential spin-down contributions. 
Additionally, we follow~\citep{O3aSNRweave, fullo3_casA} and adopt a slightly more conservative approach, allowing the upper limit on $\dot{f}$ to reach zero as follows:
\begin{align}
\label{eq:weave_ranges}
  -{\frac{f}{\tage}} \leq \dot{f} \leq 0, \quad
  2{\frac{\dot{f}^2}{f}} \leq \ddot{f} \leq 7{\frac{\dot{f}^2}{f}}.
\end{align}
Setting the upper limit on $\dot{f}$ to zero allows for time-dependent braking indices and uncertainties in the source's effective age.
Table~\ref{tab:weave target param} lists the ages of the targets used to configure the search. 
It also includes the maximum absolute values of $\dot{f}$ and $\ddot{f}$ at the lowest and highest search frequencies.

\begin{table*}[!tbh]
\begin{center}
  \begin{tabular}{lccc}\toprule
    \T\B                            &  G111.7$-$2.1/\casa  &   G266.2$-$1.2/\vela &  \targetg    \\
    \midrule
    \T\B $t_{\rm age}$ (year)             &  300              &   690 & 1600 \\
    \hline
    \T\B $\dot{f}$ range (Hz/s) @20 Hz          &  $[-\sci{2.1}{-9},\;0]$                     &   $[-\sci{9.2}{-9},\;0]$  &  $[-\sci{4.0}{-10},\;0]$ \\
    \T\B $\ddot{f}$ range (Hz/s$^2$)  @20 Hz     &  $[0,\;\sci{1.6}{-18}]$                    &   $[0,\;\sci{3.0}{-19}]$  & $[0,\;\sci{5.6}{-20}]$  \\
    \hline
    \T\B $\dot{f}$ range (Hz/s) @475 Hz         &  $[-\sci{5.0}{-8},\;0]$                     &   $[-\sci{2.2}{-8},\;0]$   & $[-\sci{9.4}{-9},\;0]$  \\
    \T\B $\ddot{f}$ range (Hz/s$^2$) @475 Hz    &  [0,\;$\sci{3.7}{-17}]$                    &   $[0,\;\sci{7.0}{-18}]$  & $[0,\;\sci{1.3}{-18}]$  \\
    \hline 
    \T\B $\Tcoh$ (day)             &  5              &   7.5 &  7.5 \\
\bottomrule
\end{tabular}
  \caption{The age used to define the Weave search range, along with the search range for $\dot{f}$, $\ddot{f}$ values at 20 and 475 Hz and $\Tcoh$ used in the initial searches.}
\label{tab:weave target param}
\end{center}
\end{table*}

The Weave program~\citep{Wette2018} creates a template grid in the parameter space for each time segment that is appropriate to computing the \Fstat\ for a coherence time $\Tcoh$ equal to the total observation period $\Tobs$ divided by $\Nseg$. 
The spacing of the grid points in ($f$, $\dot{f}$, $\ddot{f}$) is set according to a metric~\citep{Wette2013,Wette2015} that ensures a worst-case maximum mismatch ($\mcoh$) defined by the fractional loss in \Fstat\ value due to a true signal not coinciding with a search template.

Separately, a much finer grid is defined for the entire observation period with respect to the midpoint of the observation period, with its own mismatch parameter $\msemicoh$, analogous to $\mcoh$. 
This parameter, similar to $\mcoh$, is the average of the coherent mismatch values across all segments~\citep{Wette2015}. Its value is chosen based on a balance between sensitivity and computational cost.
At initialization, the Weave pipeline creates a mapping between each point in the semi-coherent template grid and the nearest point in each of the coarser segment grids, accounting for frequency evolution. The semi-coherent detection statistic $\FM$ is calculated as the average of the $\F$ values across the $\Nseg$ segments for each semi-coherent template based on this mapping:
\begin{equation}
  \FM = \frac{1}{\Nseg}\sum_{i=1}^{\Nseg}\, \F. 
\end{equation}
For the searches presented here, we use the same mismatch parameters $\mcoh=0.1$ and
$\msemicoh=0.2$ as used in \cite{O3aSNRweave}.

We divide the search jobs into 0.1-Hz bands of $f$, with further divisions in $\dot{f}$, as needed, to keep each job's computational duration less than 24 hours, for practical reasons.
Table~\ref{tab:weave fdotdivisions} shows the $\dot{f}$ divisions width for different frequency bands for the searches.
\begin{table}[tbh!]
\begin{center}
\centering
  \begin{tabular}{lcc}\toprule
     Source  & $f$ (Hz) & $\dot{f}$ width (Hz/s) \\
    \midrule
    \T\B G111.7$-$2.1/\casa       & 20--200    & $1.5\times 10^{-9}$\\
    \T\B G111.7$-$2.1/\casa      & 200--475 & $1\times 10^{-9}$  \\
    \T\B G266.2$-$1.2/\vela       & 20--475   & $8\times 10^{-10}$   \\
    \targetg  & 20--475   & $8\times 10^{-10}$ \\
    \bottomrule 
\end{tabular}
  \caption{Width of $\dot{f}$ sub-ranges for Weave's initial search jobs (0.1-Hz sub-bands) to keep job durations under 24 computational hours. Each sub-band is limited to a top-list of 1000 candidates.}
  \label{tab:weave fdotdivisions}
\end{center}
\end{table}
Each individual search job returns the ($f,\dot{f},\ddot{f}$) values of the 1000 templates (``top-list'') with the largest (``loudest'') $\FM$ in each 0.1-Hz bands with $\dot{f}$ divisions.
We sum up the number of templates for each 1-Hz band and calculate the nominal threshold
\begin{equation} \label{eq:weave threshold}
\FMthresh = \frac{1}{N_\text{seg}}  \mathrm{CDF}^{-1}_{\chi^2}\left(1 - \frac{1}{N_\text{temp}}, 4N_\text{seg}\right),
\end{equation}
where $\mathrm{CDF}^{-1}_{\chi^2}$ is the inverse cumulative distribution function (CDF) of the $\chi^2$ distribution with $4N_\text{seg}$ degrees of freedom, evaluated at $1 - 1/N_\text{temp}$ percentile, and $N_\text{temp}$ is the number of templates in a 1-Hz band.
This threshold is set based on a signal-free $\chi^2$ distribution with four degrees of freedom per segment, such that the expected number of candidates is one per 1-Hz band in Gaussian noise.
We then fit this threshold to the function $\FMthresh(f) = \FstatHz f^a$. 
The fitted parameters are listed in Table~\ref{tab:threshparameters}. 
Templates with $\FM$ values exceeding this threshold are considered candidates.
However, non-Gaussian artifacts lead to much higher candidate counts in particular bands contaminated by instrumental line sources. 
In some cases, strong instrumental lines can cause all templates returned by a single search job to have \Fstat values exceeding the nominal threshold.
This occurs for a subset of the 0.1-Hz band and range of $\dot{f}$ searched.
We refer to those cases as ``saturated'' as potentially interesting templates can be suppressed by the top-list cap.

\begin{table}[tbh!]
\begin{center}
  \begin{tabular}{lccc}\toprule
    \T\B           & $f$  (Hz)          &  $\FstatHz$  &   $a$   \\
    \midrule
    \T\B G111.7$-$2.1/\casa&      20--475    &  7.11        &  0.0186      \\
    \T\B G266.2$-$1.2/\vela   & 20--475      &  7.85        &  0.0218      \\
    \T\B \targetg    &20--475       &  7.71        &  0.0198      \\
\bottomrule 
\end{tabular}
  \caption{Parameters defining the analytic threshold function $\FMthresh(f) = \FstatHz f^a$ applied to the $\FM$ detection statistic to define initial candidates for Weave's follow-up.}
\label{tab:threshparameters}
\end{center}
\end{table}

\subsection{Postprocessing and follow-up} \label{appendix: weave follow-up}
For non-saturated sub-ranges of individual 0.1-Hz bands, candidates exceeding the threshold $\FMthresh(f)$ are followed up in a sequential procedure where at each step, the coherence time $\Tcoh$ is increased (and hence the number of segments $\Nseg$ is decreased). 
Because the non-centrality parameter for the $\F$ detection statistic scales approximately linearly with $\Tcoh$, one expects a similar percentage increase in the $\FM$. 
To determine the required ratio increase in $\FM$, We perform 50 software injections (signal simulations) with randomly chosen frequency parameters within each 1 Hz band and with $h_0=\hsens$ that yields 95\%\ detection efficiency (see Sec.~\ref{appendix: weave ul injections}).
We then follow up candidates with $\FM$ above the $\FMthresh(f)$ threshold and include higher order terms in $\Phi(t)$ to better capture the signal,
given the coherence time and frequency derivative order configuration for each follow-up stage, as shown in Table~\ref{tab:weave follow-up config}.
\begin{table*}[tbh!]
\begin{center}
\begin{tabular}{ccc}\toprule
 & G111.7$-$2.1/\casa & G266.2$-$1.2/\vela, \targetg \\
\midrule
\T\B 1st follow-up $\Tcoh$ (highest order in $\Phi(t)$) change    & $5 \; \mathrm{days}\rightarrow10\;\mathrm{days}$ $(2\rightarrow 2)$ & $7.5\;\mathrm{days}\rightarrow15\;\mathrm{days}$ $(2\rightarrow 2)$ \\
\hline
\T\B 2nd follow-up   & $10\rightarrow20$ $(2\rightarrow 2)$ & $15\rightarrow30$ $(2\rightarrow 2)$ \\
\hline
\T\B 3rd follow-up   & $20\rightarrow40$ $(2\rightarrow 3)$ & $30\rightarrow60$ $(2\rightarrow 3)$ \\
\hline
\T\B 4th follow-up  &  - & $60\rightarrow120$ $(3\rightarrow 4)$ \\
\hline
\T\B 5th follow-up& - & $120\rightarrow240$ $(4\rightarrow 4)$ \\
\bottomrule
\end{tabular}
\caption{Coherence time and frequency derivative order configuration for different Weave\ follow-up stages.}
\label{tab:weave follow-up config}
\end{center}
\end{table*}
The requirement for the increase in $\FM$ to recover the injected signals above the $\FMthresh(f)$ threshold for all follow-up stages are listed in Table~\ref{tab:mean2F ratio increase requirement}. 

\begin{table}[tbh!]
\begin{center}
\begin{tabular}{lccc}\toprule
  & G111.7$-$2.1/\casa & G266.2$-$1.2/\vela & \targetg \\
\midrule
\T\B 1      & 32\% & 36\% & 35\% \\
\T\B 2       & 41\% & 49\% & 49\% \\
\T\B 3       & 62\% & 62\% & 60\% \\
\T\B 4       & - & 68\% & 65\% \\
\T\B 5      & - & 74\% & 70\% \\
\bottomrule
\end{tabular}
  \caption{$\FM$ increase requirement in each stage of Weave\ candidate follow-up.}
  \label{tab:mean2F ratio increase requirement}
\end{center}
\end{table}

In addition, following up only the top 10 loudest survivors at later follow-up stages is sufficient to recover the injected signals with high efficiency.
Therefore, although we evaluate all candidates above the $\FMthresh(f)$ threshold from the initial search stage,
in the successive stages only the top 10 loudest survivors per initial candidate are evaluated for further follow-up. We have verified
that these threshold criteria preserve greater than 95\%\ recovery efficiency for the injections, while allowing conservatively for an uncertainty
in source right ascension and declination of $\pm10^{-5}$ radians ($\sim\pm$2 arcsec).
The loudest candidate in each saturated sub-range is followed up using the same procedure to ensure that if the saturation were due to an unusually strong signal, it would not be suppressed by the top-list cap.

We further validated the follow-up procedure with four ``hardware injections" that have been injected in the LIGO detector during the run via modulation of interferometer mirror positions via auxiliary lasers~\citep{BiwerEtal}.
We chose Injections \textit{3, 6, 8,} and \textit{12}, ranging from $\sim37$ Hz to $\sim189$ Hz, which use the same simulation parameters as used in the O3 run~\citep{O3aAllSky}, except for somewhat reduced amplitude. 
All signals were successfully recovered.

For \casa/\vela/\targetg, there were $\sim$$\sci{1.3}{5}$/$\sci{1.2}{5}$/$\sci{6.9}{4}$ candidates above the initial search threshold in bands that were not saturated (not excluded from consideration by severe instrumental artifacts).
The parameters of the surviving candidates after the second follow-up stage and the complete lists of the saturated sub-bands can be found in~\cite{cheung_2025_zenodo}.

For \casa,  repeatedly doubling the coherence time to $\Tcoh=40$ days eliminates all candidates.
For \targetg, we eliminate all candidates in the final follow-up stage when $\Tcoh=240$ days.
On the other hand, two clusters of candidates for \vela survive the final follow-up stage ($\Tcoh=240$ days) with source frequencies near $247$ Hz.
In this case, there is a prolonged period during the run when an instrumental line overlaps the nominal signal's frequency trajectory {\it vs.} time, a period during which there is a disproportionate contribution to the detection statistic. 
Hence we do not consider either candidate cluster to be a credible astrophysical signal.

We conclude that there is no significant evidence in the Weave pipeline analysis for a CW signal from the Central Compact Objects in the SNRs \casa, \vela, and \targetg.

\subsection{Upper limits} \label{appendix: weave ul injections}
\begin{figure}
  \centering
  \includegraphics[width=1\linewidth]{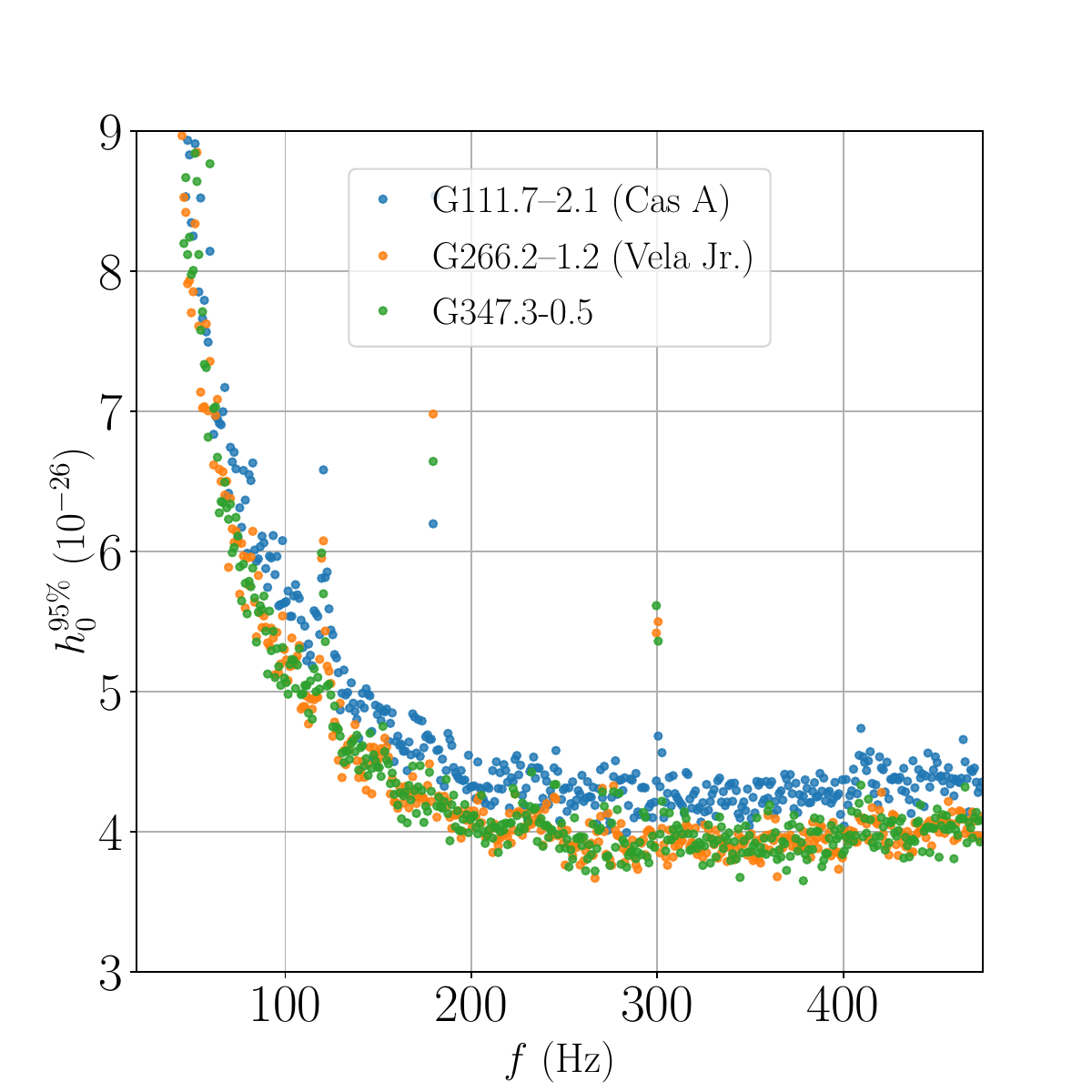}
  \caption{GW strain amplitude upper limits (95\% confidence level) in each 1-Hz band for the \casa (blue), \vela (orange), and \targetg (green) searches using Weave. 
  All the upper limits presented here are below the age-based strain amplitude limit ($> 10^{-24}$) for the three targets. 
  }
  \label{fig:weave_sensitivities}
\end{figure}

In Figure~\ref{fig:weave_sensitivities}, we report the 95\% confidence level frequentist upper limits on strain amplitude $h_0^{\rm 95\%}$ for all the targets searched by Weave.
To obtain these upper limits, we conduct software injections of signals with varying amplitudes, drawing frequencies uniformly within each 1-Hz band (excluding 0.1 Hz sub-bands containing at least one $\dot{f}$ interval with saturated candidate top-lists). 
The parameters $\cos \iota$, $\psi$, $\dot{f}$, and $\ddot{f}$ are also drawn uniformly. 
An injection is deemed detectable if its $\FM$ value, computed using Weave with the initial search setup, meets or exceeds $\FMthresh(f)$
(assuming all candidates above this threshold corresponding to a true signal would be followed up and confirmed). 
This procedure is repeated for different signal amplitudes $ h_0 $, and the detection fraction is fitted using a sigmoidal function
\begin{equation}
 p(h_0) = \frac{1}{1 + \exp\left(\frac{a - h_0}{b}\right)},
 \label{eqn:sigmoid}
\end{equation}
where $ p(h_0) $ represents the fraction of detected injections for a given $ h_0 $, and $ a $ and $ b $ are the parameters of the sigmoidal function. 
The number of injections is chosen to ensure that the statistical uncertainty in the estimated $ h_0^{\rm 95\%} $ is below 5\%.

\bibliographystyle{aasjournal}
\bibliography{O4aSNR}

\end{document}